\documentclass[preprint,3p,11pt]{elsarticle}
\usepackage{amsmath}
\usepackage{amsfonts}
\usepackage{amssymb}
\usepackage{graphicx}
\usepackage{dcolumn}
\usepackage{bm}
\usepackage[colorlinks = true, allcolors = blue]{hyperref}
\usepackage{float}
\usepackage{epstopdf}
\usepackage{lineno}

\makeatletter
\def\ps@pprintTitle{%
 \let\@oddhead\@empty
 \let\@evenhead\@empty
 \def\@oddfoot{}%
 \let\@evenfoot\@oddfoot}
\makeatother

\begin{document}
\begin{frontmatter}
	\title{Electronic transport in two-dimensional strained Dirac materials under multi-step Fermi velocity barrier: transfer matrix method for supersymmetric systems}
	
	\author[ad1,ad2]{Anh-Luan~Phan}
	\ead{phananhluan@duytan.edu.vn}
	
	\author[ad3,ad4]{Dai-Nam~Le\corref{cor1}}
	\cortext[cor1]{corresponding author}
	\ead{ledainam@tdtu.edu.vn} 
		
	\address[ad1]{Institute of Fundamental and Applied Sciences, Duy Tan University, Ho Chi Minh City, 700000, Viet Nam}
	\address[ad2]{Faculty of Natural Sciences, Duy Tan University, Da Nang City, 550000, Viet Nam}
	\address[ad3]{Atomic~Molecular~and~Optical~Physics~Research~Group, Advanced Institute of Materials Science, Ton Duc Thang University, Ho~Chi~Minh~City, Vietnam}
	\address[ad4]{Faculty of Applied Sciences, Ton Duc Thang University, Ho~Chi~Minh~City,~Vietnam}		
		
	\begin{abstract}
In recent years, graphene and other two-dimensional Dirac materials like silicene, germanene, etc. have been studied from different points of view: from mathematical physics, condensed matter physics to high energy physics. In this study, we utilize both supersymmetric quantum mechanics (SUSY-QM) and transfer matrix method (TTM) to examine electronic transport in two-dimensional Dirac materials under the influences of multi-step {deformation} as well as multi-step Fermi velocity barrier. The effects of multi-step effective mass and multi-step applied fields are also taken into account in our investigation. Results show the possibility of modulating the Klein tunneling of Dirac electron by using strain or electric field.
	\end{abstract}
\end{frontmatter}

\section{Introduction}
Graphene and its cousins such as silicene, germanene, etc. \cite{Novoselov2005, Liu2011a, Liu2011b, Wehling2014, Woods2016} have taken the lead in investigating the phenomenal analogy between different fields of physics. Amazingly, in these two dimensional ($2D$) Dirac materials, there is the (ultra)relativistic behaviour of the quasi-particles near $K$ or $K^{\prime}$ valleys, i.e their low energy propagation is governed by a Dirac-Weyl equation instead of the usual Schr\"odinger equation \cite{Zhang2005, Neto2009, Tahir2013, Tahir2013b, Matthes2013, Wang2015}. This unique property of $2D$ Dirac materials links the low energy phenomena in condensed matter physics to the high energy ones in quantum field theory. Especially, the massless quasiparticle of graphene exhibits a remarkable  phenomenon, namely the Klein tunnelling or Klein paradox \cite{Katsnelson2006} when applying a static electro voltage barrier. This ultrarelativistic quantum phenomenon \cite{Klein1929, Calogeracos1999} comes purely from the nature of massless Dirac fermion in graphene and make graphene acts like metal whereas the electric current cannot be turned off easily \cite{Neto2009}.  Therefore,  if one aims to electronic applications based on controlling the electric current, it is natural to think about trapping the Dirac fermion of graphene by using a magnetic field or finding other materials similar to graphene with a small bandgap.

The first approach has been widely examined in Dirac materials using both constant magnetic field \cite{Zhang2005, Tahir2013, Tahir2013b, Tabert2013} and inhomogeneous magnetic fields \cite{Kuru2009, Masir2011, Roy2012, Downing2016, LE2020PhysE}. In this case, the Dirac electron of graphene and other Dirac materials will be bound in Landau states by a magnetic field and forms the Landau levels which are proportional to $\sqrt{n}$ instead of $n$ like two-dimensional electron gas \cite{Goerbig2011, Zhang2005, Kuru2009, Masir2011, Roy2012, Tahir2013, Tahir2013b, Tabert2013, Downing2016, LE2020PhysE}. This property is a fingerprint for the relativistic behaviour of Dirac electron. Consequently, using (multi) magnetic barrier, we may efficiently control the transport property, i.e the tunnelling  of Dirac electron \cite{Masir2008, DellAnna2009}. On the other hand, to open the bandgap for honeycomb lattice monolayer, two carbon atoms in a unit cell can be replaced with two atoms from different elements to break its inverse symmetry, for example, hexagonal boron nitride monolayer \cite{Wang2015}. Another way to achieve graphene-like materials with small band gap is using two identical atoms from other elements with stronger spin-orbit coupling such as silic (Si), germanium (Ge) or gray tin ($\alpha$-Sn) \cite{Liu2011a, Liu2011b, Wehling2014, Wang2015, Nguyen2021} to build monolayers of silicene, germanene, and tinene. Besides their small band gaps, these monolayers have a remarkable property that their bandgap can be tuned electronically by applying a voltage on bulk direction \cite{Drummond2012}. In the effective mass perspective, this means their Dirac fermion has a small tunable mass \cite{Matthes2013, Tahir2013, Tahir2013b, LE2020PhysE}. The effect of this tunable mass on Landau levels of these Dirac materials has also been examined in several works such as References \cite{LE2020PhysE, LE2020167473} and references therein.

Apart from these two approaches, nowadays, strain engineering is the most contemporary way to manipulate the electronic property of two-dimensional Dirac materials. This approach is based on two different effects from deforming $2D$ honeycomb lattices known as strain-induced pseudogauge field and modulated Fermi velocity \cite{Vozmediano2008, Vozmediano2010, Pereira2009, Low2010, Levy2010, Guinea2010, DeJuan2013, Raoux2010, Pellegrino2011, DeJuan2012, Salvador2013, Sloan2013, Sanjuan2014, Oliva2015, Downing2017, Naumis2017}. Notably, unlike the real gauge field of an electromagnetic field, the strain-induced pseudogauge field is valley-dependent. Particularly, the effect of strain-induced pseudomagnetic field in $K^{\prime}$ valley is opposite to the one in $K$ valley. This unique property of strain-induced pseudogauge field provides the possibility of valleytronics. As a result of pseudomagnetic field induced by strain, Dirac electron in graphene is also confined into bound states to form the Landau levels and consequently reveals some quantum phenomena such as quantum Hall effect, de Haas - van Alphen \cite{Vozmediano2008, Vozmediano2010, Pereira2009, Low2010, Levy2010, Guinea2010,  DeJuan2013, Lantagne2020, Ma2020} or Landau contraction when applying in-plane electric voltage \cite{Le2020EPJB}. Combining strain-induced pseudogauge field and magnetic barrier could provide a good valley filter \cite{Zhai2010}. Beside pseudogauge field, deforming $2D$ honeycomb lattice can produce the spatially dependent \cite{Pellegrino2011, DeJuan2012, Oliva2015, Naumis2017} or anisotropic \cite{Oliva2015, Naumis2017, ocampo2020} Fermi velocity of the Dirac electron. Anisotropic Fermi velocity of strained graphene breaks the equivalence between the armchair and zigzag directions and thus significantly affects on Dirac electron under the presence of an in-plane electric field. For example, Reference \cite{LE2020167473} has examined the effect of anisotropic Fermi velocity on the Landau contraction and de Haas - van Alphen oscillation of magnetized graphene sheet under the presence of an electric field. The non-uniform strain which induced spatially dependent Fermi velocity can help us localize the Dirac electron by forming bound states \cite{Downing2017, Ghosh2017, Phan2020a, Ghosh2021}. Similar to the pseudomagnetic field, the position-dependent Fermi velocity also manipulates transport property of Dirac electron in magnetic field \cite{Oliva2020} or magnetic superlattice \cite{BEZERRA2020}. 

In view of the above observations, we believe it is relevant to observe the influence of multi-barrier Fermi velocity as well as multi-barrier rest energy on the electronic transport property of Dirac electron in several $2D$ Dirac materials. This could give us a chance to manipulate the Klein tunnelling of Dirac electron by using strain or electric field. To do so, we examine the problem of $2D$ Dirac electron with Fermi velocity as well as its rest energy are both multi-step barrier profile under the presence of multi-barrier in-plane electric voltage combined with a multi-barrier magnetic vector field. From  the experimental point of view, the considered system can be built by putting a monolayer of $2D$ Dirac material inside an electric cavity as well as setting up a series of in-plane electrodes to provide its Dirac fermion a position-dependent mass and a scalar potential. To manipulate the Fermi velocity of the Dirac fermion, a suitable deformation can be a good choice. Then the system should be placed near superconducting materials to create a position-dependent magnetic field. This setup makes our system look like a lens or a Veselago lens in single-barrier case or a Fabry–-P{\'e}rot cavity in a multi-barrier case. The details of our system will be described in the next Section. By considering a suitable ansatz of the applied fields as in Reference \cite{Phan2020a}, we exploit the supersymmetry (SUSY) nature of our system to get the exact expression of wavefunction versus energy for step profiles. Then our results will be generalized into multi-barrier profiles via the transfer matrix method (TTM), see Reference \cite{Wang2010} for example. Finally, observing the profile of transmission probability will give us the answer of how multi-barrier Fermi velocity, as well as multi-barrier rest energy, affects Klein tunneling of the Dirac materials.      

After this section, our paper is presented as follows. Section \ref{sec:2} introduces our considered system as well as the formalism of supersymmetric quantum mechanics to find out wavefunction versus the energy of Dirac materials. Next, the solution for step profiles is provided in Section \ref{sec:3}. A semi-classical interpretation is also given to explain our observations. Then the transfer matrix method in the scheme of supersymmetric quantum mechanics is utilized for multi-step barrier profile of effective mass, scalar, and vector potentials in Section \ref{sec:4}. Also, the profile of transmission probability is examined in this section. Section \ref{sec:5} discusses the Klein tunnelling and how multi-barrier Fermi velocity, as well as multi-barrier rest energy, has its impact on Klein tunnelling of the Dirac materials while Section \ref{sec:6} sums up our conclusion.

\section{\label{sec:2} Two-dimensional Dirac materials under barriers of strain, Fermi velocity and gap}

We start with a flat sheet of $2D$ Dirac material, in which the Fermi velocity and the band gap are both position-dependent, that is, $v(\vec{r})$ and $\Delta(\vec{r})$. The sheet is exposed into a static electromagnetic field characterized by the vector and scalar potentials $\vec{A}(\vec{r})$ and $\phi (\vec{r})$. Here we are interested in the profiles of Fermi velocity, bandgap, magnetic and electric fields which vary along $x$-axis, i.e. $v(\vec{r}) = v(x)$, $\Delta(\vec{r})=\Delta(x)$, $\vec{A}(\vec{r}) = (0,A_y(x),0)$ (Landau gauge) and $\phi(\vec{r}) = \phi(x)$.
The proper Dirac Hamiltonian describing such a system is \footnote{All the calculations in this work will be carried out with the use of the following dimensionless units because of their convenience: the lattice constant $a_0$, $v_F$, $B_0= \hbar e^{-1} a_0^{-2}$, $E_0 =\hbar v_F e^{-1} a_0^{-2}$, $\varepsilon_0 = \hbar v_F a_0^{-1}$ as units of length, velocity, magnetic field strength, electric field strength and energy, respectively. Formally, we can set $e=\hbar=v_F=1$.} \cite{Peres2009, Phan2020a, Ghosh2021}
\begin{eqnarray}\label{H}
\hat{H} = v(x) \left[\sigma_x (-i \partial_x + \tilde{A}(x)) + \sigma_y (-i \partial_y + A_y(x))\right] + \sigma_z \Delta(x)/2 + \phi(x),
\end{eqnarray}
where $\sigma_{x,y,z}$ are the Pauli matrices, $\vec{\hat{p}} = (\hat{p}_x, \hat{p}_y) = (-i \partial_x, -i \partial_y)$ and the fictitious vector potential
\begin{equation}
\tilde{A}(x) = - i \dfrac{\partial_x v(x)}{2 v(x)},
\end{equation}
which arises from the position-dependence of the Fermi velocity to retain the Hermiticity of the Hamiltonian. In this work, effective bandgap $\Delta(x)$ is controlled by an perpendicular electric field $\Delta (x) = \Delta_0 + \kappa \mathcal{E}_z$ (see in Reference \cite{LE2020PhysE} and reference therein for value of $\kappa$ in different Dirac materials) while the in-plane biaxial deformation
\begin{equation}
    u_x = \int f(x) dx, \quad u_y = f(x) y, \quad u_z = 0,
\end{equation}
whose strain tensor is
\begin{equation}
    \epsilon = \left[ \begin{matrix} \partial _x u_x & (\partial _y u_x + \partial _x u_y)/2 \\ (\partial _y u_x + \partial _x u_y)/2 & \partial _y u_y \end{matrix} \right] = \left[ \begin{matrix} f(x) & y f'(x)/2 \\
  y f'(x)/2 & f(x)
  \end{matrix} \right],
\end{equation}
is used to produce the position-dependent Fermi velocity. Particularly, in this work, we only consider the step profile of $f(x)$, thus $f'(x) = 0$ i.e $\epsilon = f(x) \mathbf{1}_{2\times 2}$ and then the position-dependent Fermi velocity is simply become  \cite{Vozmediano2010, Oliva2015}:
\begin{equation}
      v (x) = v_F \left[ 1 + (1- \beta) f(x) \right], 
\end{equation}
whereas $\beta = 2$ to $3$ is  Gr{\"u}neisen parameter and $v_F$ is Fermi velocity of pristine graphene monolayer. Noticeably, the position-dependent Fermi velocity always positive $v(x) > 0$; thus, the strain profile $f(x)$ must be less than $1/(\beta - 1)$. This condition is always satisfied since continuum approximation for Dirac materials is only suitable for low-strength deformation.

We need to solve the following Dirac equation
\begin{equation}\label{originalDirac}
H \Psi(x,y) = \varepsilon \Psi(x,y),
\end{equation} 
with the Dirac Hamiltonian \eqref{H}. From the previous results \cite{Phan2020a}, at this point it is straightforward to transform Equation \eqref{originalDirac} into a $(1+1)$-dimensional Dirac equation for the intermediate pseudo-spinor $K(w)$
\begin{eqnarray}\label{K-eqn}
\left\{\sigma_x (-i \partial_w) + \sigma_y v(w) (k +  A_y(w)) + \sigma_z \Delta(w)/2 + \phi(w) - \varepsilon \right\} K(w) = 0
\end{eqnarray}
by using suitable transformation and variable-changing
\begin{eqnarray}
\Psi(x,y) = \dfrac{e^{i k y}}{\sqrt{v(x)}} K(x), \qquad\qquad
x \to w(x) = \int^x_0 \dfrac{du}{v(u)}.
\end{eqnarray}
Here the $e^{i k y}$ pre-factor represents the translation symmetry along $y$ axis, while the fictitious gauge field $\tilde{A}$ is removed by the factor $v^{-1/2}(x)$.

The present form of Equation \eqref{K-eqn} makes it difficult for further insights. On the other hand, our emphasis in this work is on using the SUSY formalism to the considered physical problem as well as deducing the essential properties, rather than the specific results themselves. Therefore, we choose to work with the following ansatz: 
\begin{eqnarray}\label{ansatz}
&&v(w) = q_v p(w), \qquad	A_y (w) = q_A / p(w) + A_0, \nonumber \\
&&\Delta(w)/2 = q_{m} p(w), \qquad	\phi(w) = q_\phi p(w) + \phi_0,
\end{eqnarray}
where the constants $q$'s represent the strength of the corresponding quantities while the position-dependence of the system is embedded into the function $p(w)$. Besides, the constants
\begin{eqnarray}
A_0 = - \lim_{x \to - \infty} q_A/p(w(x)), \qquad
\phi_0 = - \lim_{x \to - \infty} q_\phi p(w(x))
\end{eqnarray}
will be deliberately chosen so that the vector and scalar potentials vanish when $x \to - \infty$. Then, Equation \eqref{K-eqn} becomes
\begin{eqnarray}\label{K-eqn-2}
\left\{-i \sigma_x \partial_w + \left[ \sigma_y (k+A_0)q_v + \sigma_z q_m + \mathbb{I} q_\phi \right] p(w) + \sigma_y  q_v q_A + \phi_0 - \varepsilon \right\} K(w) = 0.
\end{eqnarray}

As shown in previous works \cite{Lukose2007, Le2020EPJB, LE2020167473, Phan2020a}, at this stage we may be able to set our consideration under a new frame of reference, in which the  effective magnetic field does not vanish (hence the index $B$, or $B$-case)
\begin{eqnarray}\label{B-eqn}
	\left[ -i \sigma_x \partial_w + \sigma_y \tilde{\mathcal{W}}(w) \right] G_B(w) &=& \tilde{\varepsilon}_B G_B(w),
\end{eqnarray}
by applying a suitable rotation $K(w) = U_B G_B(w)$\footnote{We put the expression of $U_B$ (and also $U_E$, see in the text) in \ref{rotation} to avoid making the main argument lengthy.}. Then, it is straightforward to prove the SUSY nature \cite{Cooper1995, YEKKEN2013, Concha2018, Le2020JPCM, Phan2020a, Bagchi2021} of Equation \eqref{B-eqn} with the superpotential and the eigenvalue
\begin{eqnarray}
    \tilde{\mathcal{W}}(w) &=& \sqrt{-Q}~ p(w) + \dfrac{(k+A_0) q_v^2 q_A + q_\phi (\varepsilon-\phi_0)}{\sqrt{-Q}}, \nonumber\\
    \tilde{\varepsilon}_B &=& \sqrt{\dfrac{\left[(k+A_0) q_v^2 q_A + q_\phi (\varepsilon-\phi_0)\right]^2}{-Q} - (q_v q_A)^2 + (\varepsilon-\phi_0)^2},
\end{eqnarray}
where $Q = q_\phi^2 - q_m^2 - (k+A_0)^2q_v^2$ is the determinant of $\left[ \sigma_y (k+A_0)q_v + \sigma_z q_m + \mathbb{I} q_\phi \right]$. Clearly, the above argument is physically reasonable only when $Q<0$ so that $\tilde{\mathcal{W}}(w)$ is real (and therefore bound states may be allowed to exist, as analyzed in \cite{Phan2020a}). Noticeably when there are no mass as well as position-dependent Fermi velocity i.e $w = x, p(w) = 1, q_v = 1, q_m = 0$ the rotation matrix $U_B$ coincides to the gauge transformation of Lorentz boost in References \cite{Lukose2007, Le2020EPJB, LE2020167473}. Hence, this is suggested that rotating the pseudospinor by matrix $U_B$ is a generalization of \emph{Lorentz boost}.

For $Q>0$, we have an imaginary superpotential, which suggests us to perform a \emph{complex Lorentz boost} to change the Dirac equation with imaginary magnetic field into one with real electric field \cite{Tan2010}. In fact, when $Q>0$ we can use another rotation $K(w) = U_E G_E(w)$ to transform Equation \eqref{K-eqn-2} into a Dirac equation with real electric field (hence the index $E$, or $E$-case)
\begin{eqnarray}\label{E-eqn}
	\left[ -i \sigma_x \partial_w + \sigma_y \tilde{k}_E + \tilde{\phi}(w) \right] G_E(w) = \tilde{\varepsilon}_E G_E(w).
\end{eqnarray}
where
\begin{eqnarray}
\tilde{\varepsilon}_E &=& - \dfrac{(k+A_0) q_v^2 q_A + q_\phi (\varepsilon-\phi_0)}{\sqrt{Q}}, \nonumber\\
\tilde{k} &=& i \sqrt{\dfrac{\left[(k+A_0) q_v^2 q_A + q_\phi (\varepsilon-\phi_0)\right]^2}{-Q} - (q_v q_A)^2 + (\varepsilon-\phi_0)^2}, \nonumber\\
\tilde{\phi}(w) &=& - \sqrt{Q}~ p(w).
\end{eqnarray}
Note that, with a given system and a fixed energy $\varepsilon$, whether the quasi-particle behaves as in $B$-case or $E$-case depends on its $y$-momentum $k$. Introducing
\begin{eqnarray}
k_{0,\pm} = \Theta(|q_\phi|-|q_m|) \left( \pm \sqrt{ \dfrac{q_\phi^2 - q_m^2}{q_v^2} } - A_0 \right)
\end{eqnarray}
where $\Theta(t)$ is the Heaviside Theta function, we can infer that when $k_{0,-} < k < k_{0,+}$ we will have $E$-case, otherwise $B$-case.

In the next section, we will utilize the above formulation for the problem of step profiles. Besides the wave function, the final aim is to derive the distribution of the transmission probability in terms of the incident angle.

\section{\label{sec:3}Step profiles}

First we consider the following profiles
\begin{eqnarray}
&&v(x) = r_v \left[ 1 + h \Theta (x) \right],\quad
A_y(x) = r_A / \left[ 1 + h \Theta (x) \right] - r_A,\nonumber\\
&&\Delta(x)/2 = r_m \left[ 1 + h \Theta (x) \right],\quad
\phi(x) = r_\phi h \Theta (x),
\end{eqnarray}
where $r_v>0,~ r_m, h \geq 0$ and $\Theta (x)$ is the Heaviside Theta function. This profile corresponds to the step profile of in-plane strain:
\begin{equation}
    (u_x, u_y) = \left\{ \begin{matrix} (0,0) & \text{when } & x < 0\\ (x,y) & \text{otherwise} \end{matrix} \right.. 
\end{equation}

Switching to the auxiliary variable
\begin{eqnarray}
w = \int_{0}^{x} \dfrac{\text{d}u}{r_v \left[ 1 + h \Theta(u) \right]}
=
\begin{cases}
\dfrac{x}{r_v} \text{ if } x<0,\\
\dfrac{x}{r_v (1+h)} \text{ if } x>0
\end{cases},
\end{eqnarray}
we rewrite the above quantities in terms of $w$ to show that the ansatz \eqref{ansatz} is satisfied
\begin{eqnarray}
&&v(w) = q_v p(w),\quad
A_y(w) = q_A / p(w) + q_A,\nonumber\\
&&\Delta(w)/2 = q_m p(w),\quad
\phi(w) = q_\phi p(w) + q_\phi,
\end{eqnarray}
where
\begin{eqnarray}
q_v = -r_v < 0, \quad q_A = -r_A, \quad q_m = -r_m \le 0, \quad q_\phi = - r_\phi, \quad p(w) = - \left[ 1 + h \Theta(w) \right] < 0.
\end{eqnarray}

\subsection{$B$-case}

The superpotential becomes
\begin{eqnarray}
\tilde{\mathcal{W}} (w) =
\begin{cases}
- \sqrt{-Q} + \dfrac{-(k-r_A) r_v^2 r_A - r_\phi (\varepsilon+r_\phi)}{\sqrt{-Q}} = \mathcal{W}_l \qquad\qquad \text{ if $w<0$},\\
- \sqrt{-Q} (1+h) + \dfrac{-(k-r_A) r_v^2 r_A - r_\phi (\varepsilon+r_\phi)}{\sqrt{-Q}} = \mathcal{W}_r \quad\text{ if $w>0$}.
\end{cases}
\end{eqnarray}

\begin{figure}[h!]
	\centering
	\hfill
	\begin{minipage}[c]{0.45 \textwidth}
		\includegraphics[width=\linewidth]{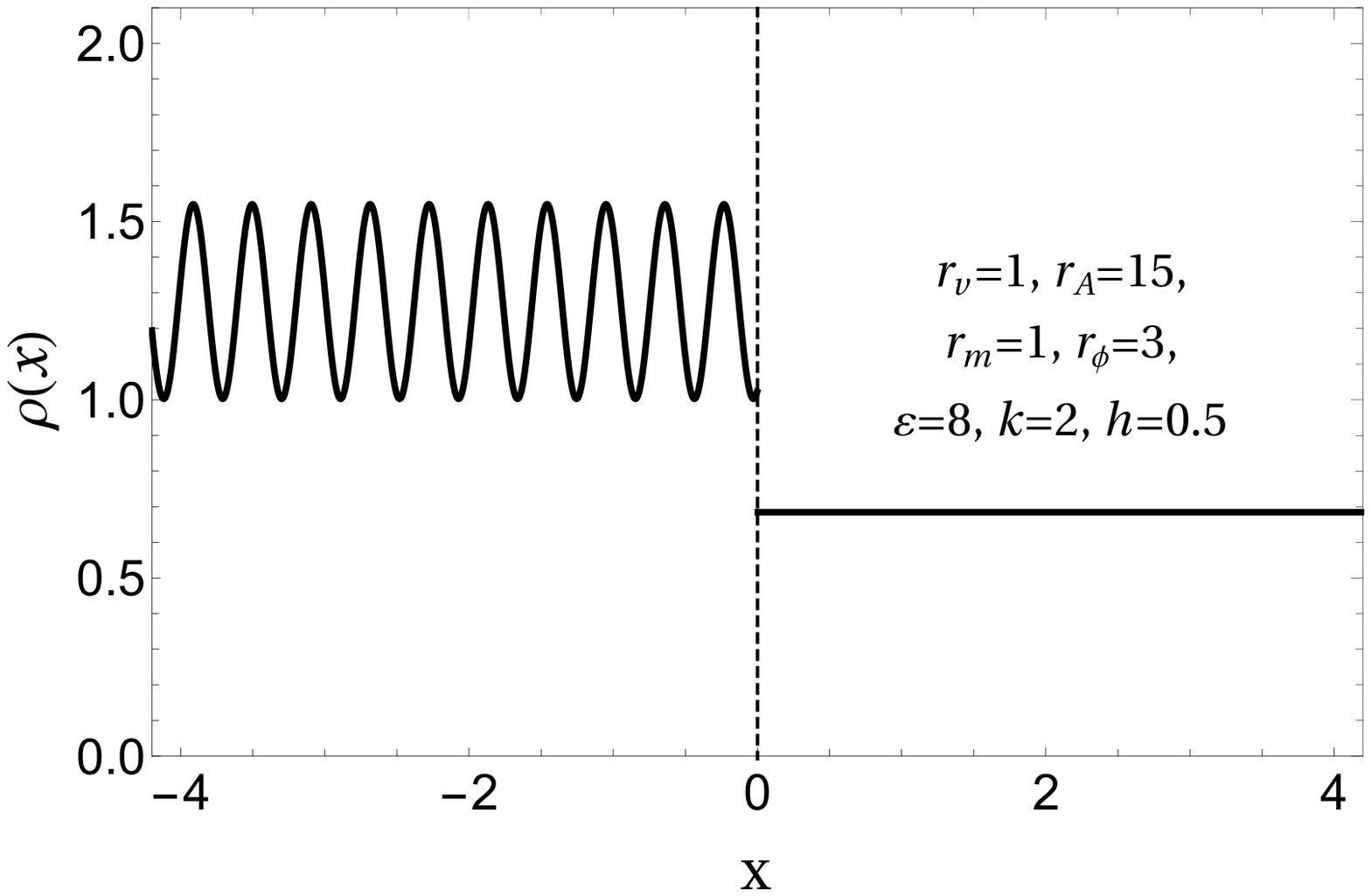}
	\end{minipage}
	\hfill
	\begin{minipage}[c]{0.38 \textwidth}
		\includegraphics[width=\linewidth]{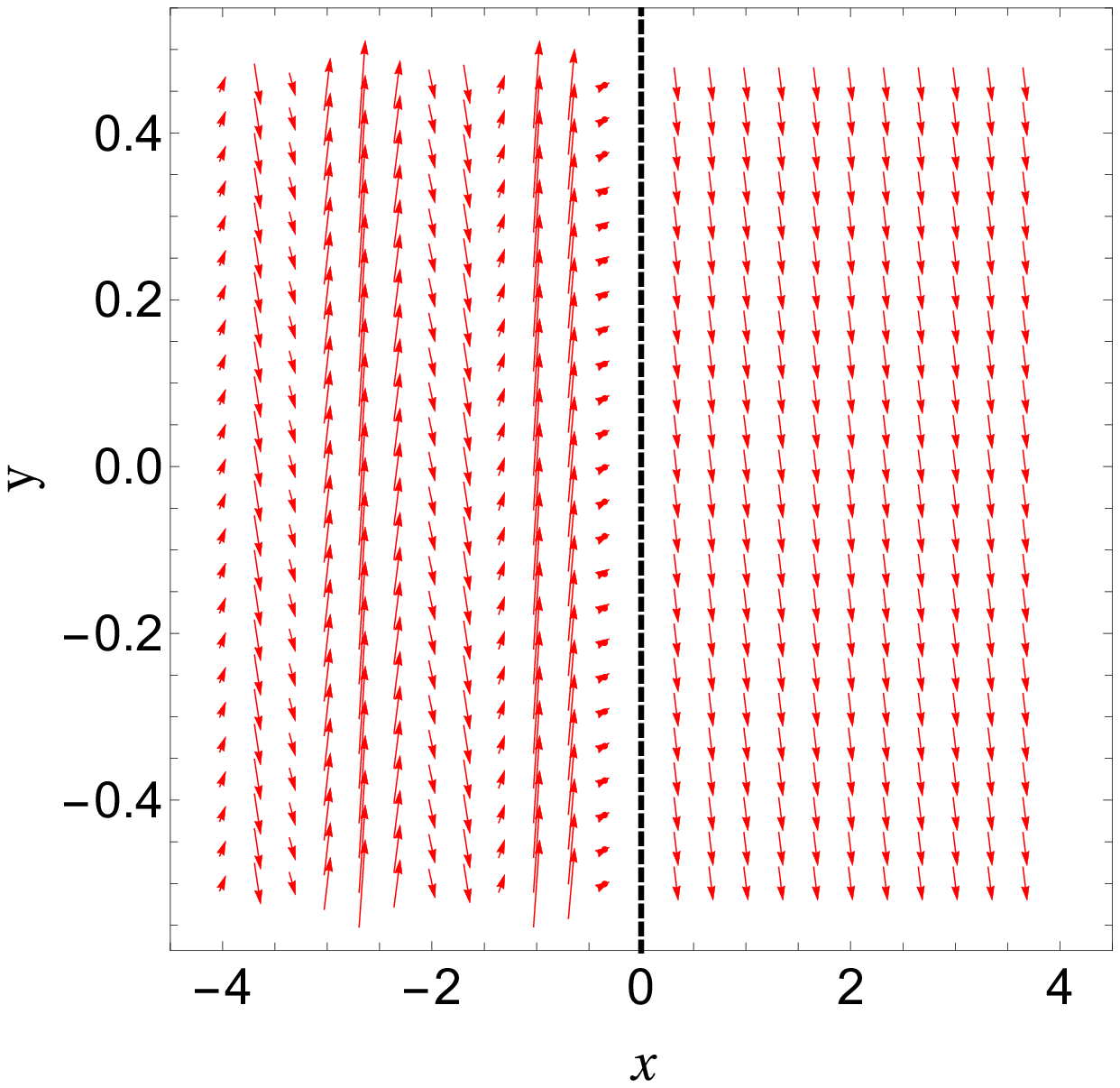}
	\end{minipage}
\hfill
	\caption{\label{rho&J-step-mag}(Color online)The probability density $\rho$ (left) and the vector plot of the corresponding probability current density $\vec{J}$ (right) of the quasi-particle for $B$-case for step profiles. The two zones are distinguished by the vertical dashed line.}
\end{figure}

By decoupling the Equation \eqref{B-eqn} into two coupled second-order differential equations, we can easily write down the scattering intermediate pseudo-spinor for a quasi-particle traveling from the left of the material sheet (i.e. $\tilde{\varepsilon}_B^2 > \mathcal{W}_g^2$)
\begin{eqnarray}
G_B(w) =
\begin{cases}
\begin{pmatrix}
c_1 e^{i q_1 w} + d_1 e^{-i q_1 w} \\
-i e^{i \theta_1^B} c_1 e^{i q_1 w} + i e^{- i \theta_1^B} d_1 e^{-i q_1 w}
\end{pmatrix} \text{ if } w<0,\\
\begin{pmatrix}
c_2 e^{i q_2 w}\\
-i e^{i \theta_2^B} c_2 e^{i q_2 w}
\end{pmatrix} \text{ if } w>0,
\end{cases}
\end{eqnarray}
where
\begin{eqnarray}\label{q1q2-mag}
q_1 = \sqrt{\tilde{\varepsilon}_B^2 - \mathcal{W}_l^2} > 0,~~
q_2 = \sqrt{\tilde{\varepsilon}_B^2 - \mathcal{W}_r^2},~~
\tan \theta_1^B = \mathcal{W}_l/q_1, ~~ \tan \theta_2^B = \mathcal{W}_r/q_2.
\end{eqnarray}
Here the coefficient $c_1$ is chosen so that the total incident current density is normalized to unity: $J_{in} = \sqrt{J_{in,x}^2 + J_{in,y}^2} = 1$, while the remaining coefficients are determined from the matching condition, being scaled by $c_1$ as
\begin{eqnarray}
\dfrac{d_1}{c_1} = \frac{2 q_1}{q_1 + q_2 - i h \sqrt{Q}} - 1, \qquad
\dfrac{c_2}{c_1} = \frac{2 q_1}{q_1 + q_2 - i h \sqrt{Q}}~.
\end{eqnarray}
The probability density $\rho(x)$ and the current density vector $\vec{J}$ are shown in Figure \ref{rho&J-step-mag}\footnote{The formula to calculate these quantities are given in \ref{rho&J}.}. A notable point is that the original pseudo-spinor $\Psi(x,y)$ does not preserve its continuity at the boundary, which is reflected by the discontinuity of the probability density $\rho(x)$. This discontinuity is due to the abrupt change in profile of the Fermi velocity, as pointed out in \cite{Peres2009}.

\subsection{$E$-case}

In this case, we also have the effective step scalar potential
\begin{eqnarray}
	\tilde{\phi} (w) = \sqrt{Q} \left[ 1 + h \Theta(w) \right].
\end{eqnarray}
The scattering solution of Equation \eqref{E-eqn} is then
\begin{eqnarray}
G_E(w) =
\begin{cases}
\begin{pmatrix}
c_1 e^{i q_1 w} + d_1 e^{-i q_1 w} \\
e^{i \theta_1^E} c_1 e^{i q_1 w} - e^{- i \theta_1^E} d_1 e^{-i q_1 w}
\end{pmatrix} \text{ if } w<0,\\
\begin{pmatrix}
c_2 e^{i q_2 w}\\
e^{i \theta_2^E} c_2 e^{i q_2 w}
\end{pmatrix} \text{ if } w>0,
\end{cases}
\end{eqnarray}
where now $q_1$, $q_2$, $\theta_1$ and $\theta_2$ are given by
\begin{eqnarray}\label{q1q2-ele}
q_1 = \sqrt{\left(\tilde{\varepsilon}_E - \sqrt{Q}\right)^2 - \tilde{k}^2} > 0, & \tan \theta_1^E = \tilde{k}/q_1, \nonumber\\
q_2 = \sqrt{\left(\tilde{\varepsilon}_E - \sqrt{Q}(1+h)\right)^2 - \tilde{k}^2}, & \tan \theta_2^E = \tilde{k}/q_2.
\end{eqnarray}
We would like to emphasize that the expressions of $q_{1}$ and $q_2$ here are the same as those in Equation \eqref{q1q2-mag}. This fact is not surprising because $q_1$ and $q_2$ actually relate to the $x$-component of the total momentum of the quasi-particle in each zone. Because the expressions of these total momenta remain unchanged in either cases (as will be shown shortly below), so do $q_1$ and $q_2$.

\begin{figure}[h!]
	\centering
	\hfill
	\begin{minipage}[c]{0.45 \textwidth}
		\includegraphics[width=\linewidth]{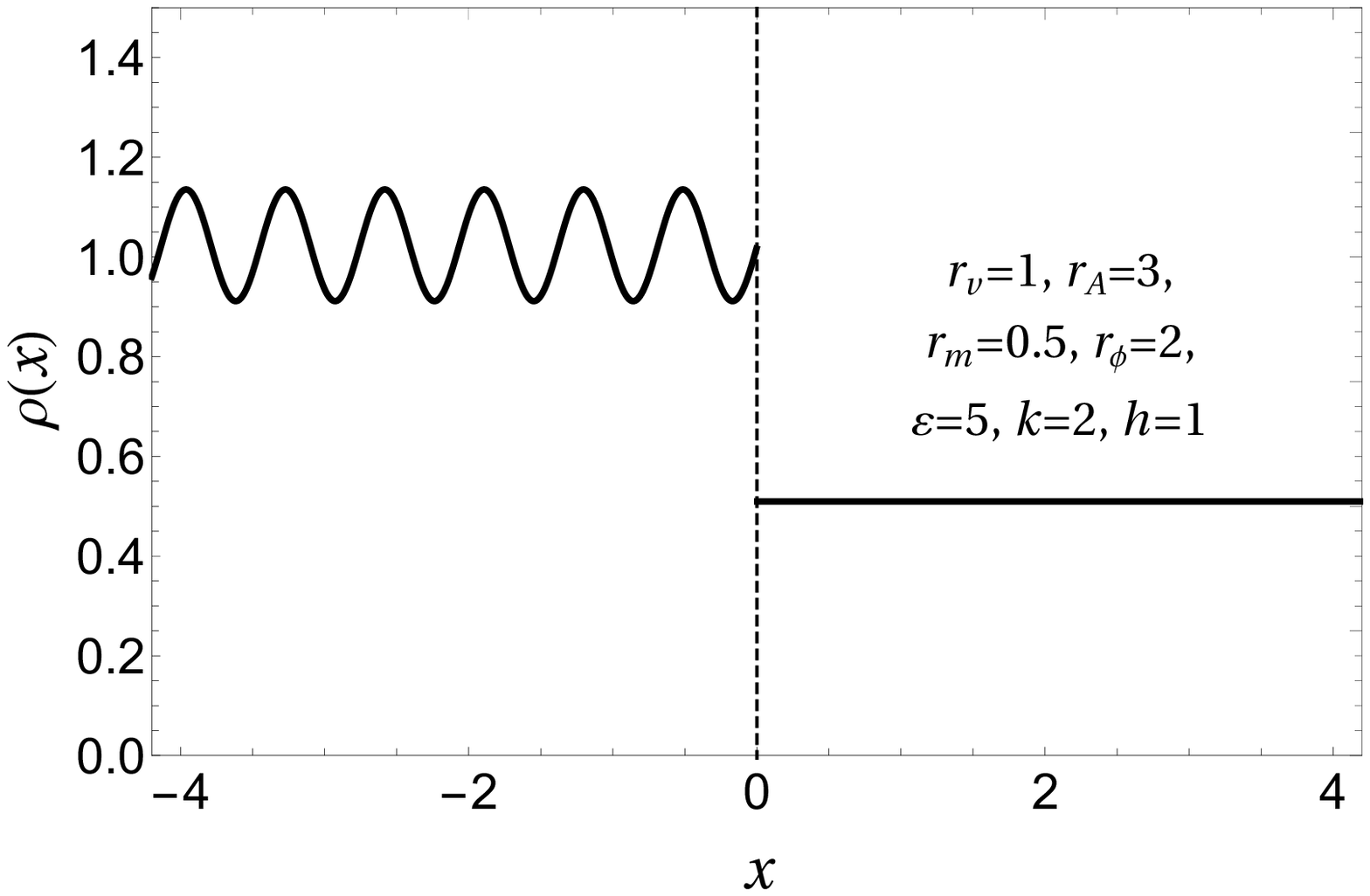}
	\end{minipage}
	\hfill
	\begin{minipage}[c]{0.38 \textwidth}
		\includegraphics[width=\linewidth]{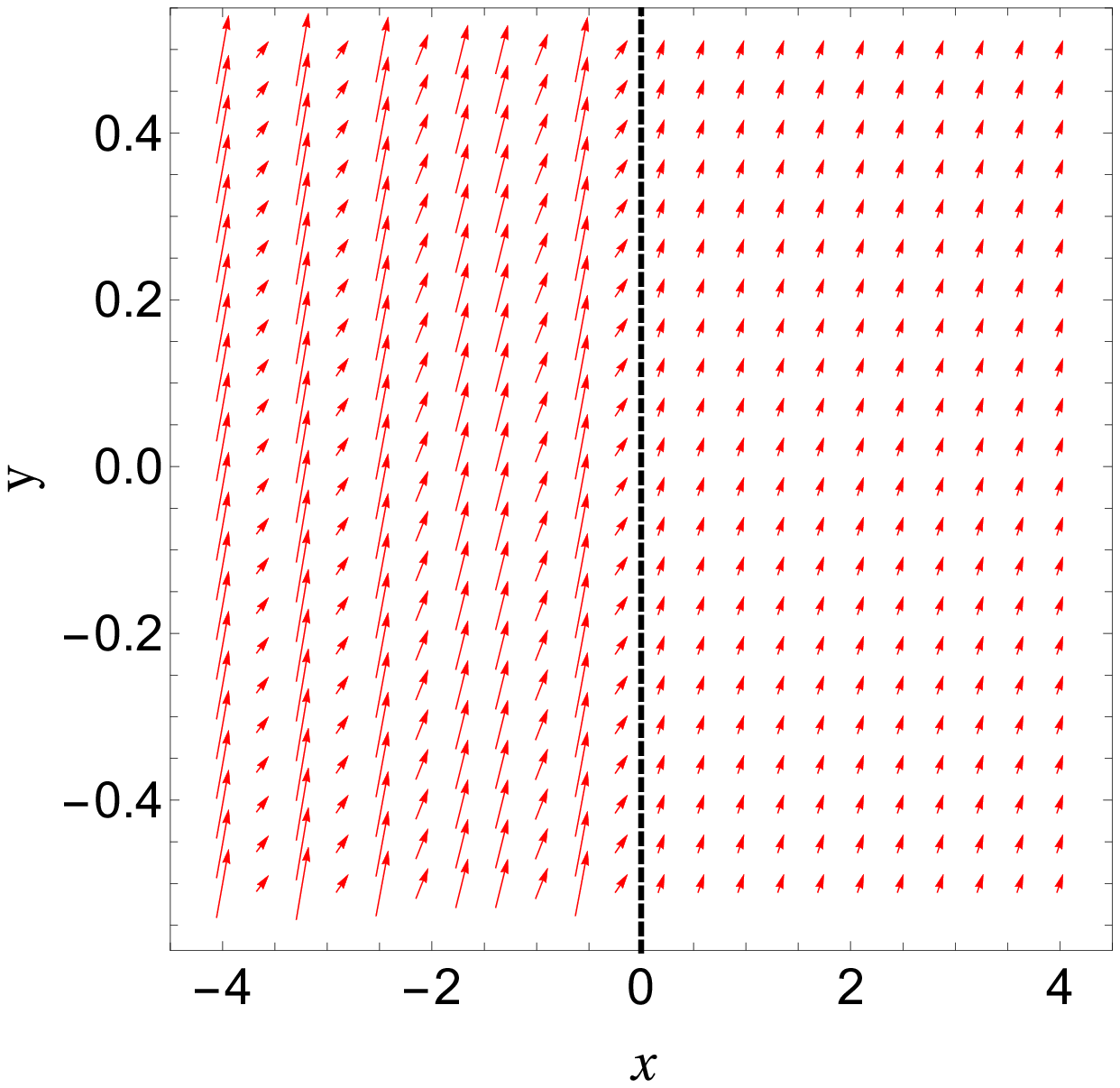}
	\end{minipage}
	\hfill
	\caption{\label{rho&J-step-ele}(Color online) The probability density $\rho$ (left) and the vector plot of the corresponding probability current density $\vec{J}$ (right) of the quasi-particle for $E$-case in step profiles. The two zones are distinguished by the vertical dashed line.}
\end{figure}

This time, the coefficients $d_1$ and $c_2$ are given by
\begin{eqnarray}
	\dfrac{d_1}{c_1} &=& \left[ \tilde{\mathcal{E}} (q_1 - q_2) - \sqrt{|Q|} \left( q_1 - q_2 + h (q_1 + i \tilde{k}) \right) \right] / D, \nonumber\\
	\dfrac{c_2}{c_1} &=& 2 q_1 \left[ \tilde{\mathcal{E}} - \sqrt{|Q|} (1+h) \right]/D, \nonumber\\
	D &=& \left[ \tilde{\mathcal{E}} (q_1 + q_2) - \sqrt{|Q|} \left( q_1 + q_2 + h (q_1 - i \tilde{k}) \right) \right].
\end{eqnarray}
The probability density $\rho(x)$ and the current density vector $\vec{J}$ are shown in Figure \ref{rho&J-step-ele}.

\subsection{The Snell-like relation}

From the above analysis, we can see that $q_1/r_v$ and $q_2/r_v(1+h)$ play the roles of $x$-momenta in the left and right zones, respectively. Thus, we deduce the case-independent total momentum in each zone
\begin{eqnarray}
	|\vec{P}| &=& \sqrt{k^2 + \dfrac{q_1^2}{r_v^2}} = \sqrt{\dfrac{\varepsilon^2 - r_m^2}{r_v^2}}, \nonumber\\
	|\vec{Q}| &=& \sqrt{\left(k - \dfrac{h r_A}{1 + h}\right)^2 + \left( \dfrac{q_2}{r_v (1+h)} \right)^2} = \sqrt{\dfrac{(\varepsilon - h r_\phi)^2}{r_v^2 (1+h)^2} - \dfrac{r_m^2}{r_v^2}}.
\end{eqnarray}
It is now natural to define the incident and refracted angles with respect to the $x$-axis as follows:
\begin{eqnarray}
	\sin \alpha = k / |\vec{P}|, \qquad
	\sin \beta = \left( k - \dfrac{h r_A}{1+h} \right) /|\vec{Q}|.
\end{eqnarray}

Here we consider only the quasi-particles whose energies satisfy the conditions $|\varepsilon| > r_m$ and $|\varepsilon - h r_\phi| > r_m (1+h)$ so that $|\vec{P}|$ and $|\vec{Q}|$ is definitely real and positive.
Keeping in mind that $k$ is a quantum number of the quasi-particle, we can deduce a Snell-like relation
\begin{eqnarray}
	&&|\vec{P}| \sin \alpha = |\vec{Q}| \sin \beta + \dfrac{h r_A}{1 + h} = k \nonumber\\
	\Leftrightarrow &&\sqrt{\dfrac{\varepsilon^2 - r_m^2}{r_v^2}} \sin \alpha = \sqrt{\dfrac{(\varepsilon - h r_\phi)^2}{r_v^2 (1+h)^2} - \dfrac{r_m^2}{r_v^2}} \sin \beta + \dfrac{h r_A}{1 + h}.
\end{eqnarray}

In this relation, $|\vec{P}|$ and $|\vec{Q}|$ play the roles of the indices of refraction in two zones. This is a generalized version in the sense that it takes many quantities of the physical system (Fermi velocity, rest mass, scalar, and vector potentials) into account. For comparison, when $r_m=r_\phi=r_A=0$, we regain the version mentioned in Equation (8) of \cite{Raoux2010} for only velocity barrier; or when  $r_m=0$, we can deduce the Equation (10) of \cite{Liu2012}.

\begin{figure}[H]
\begin{center}
	\begin{minipage}[c]{0.4 \textwidth}
		\includegraphics[width=\linewidth]{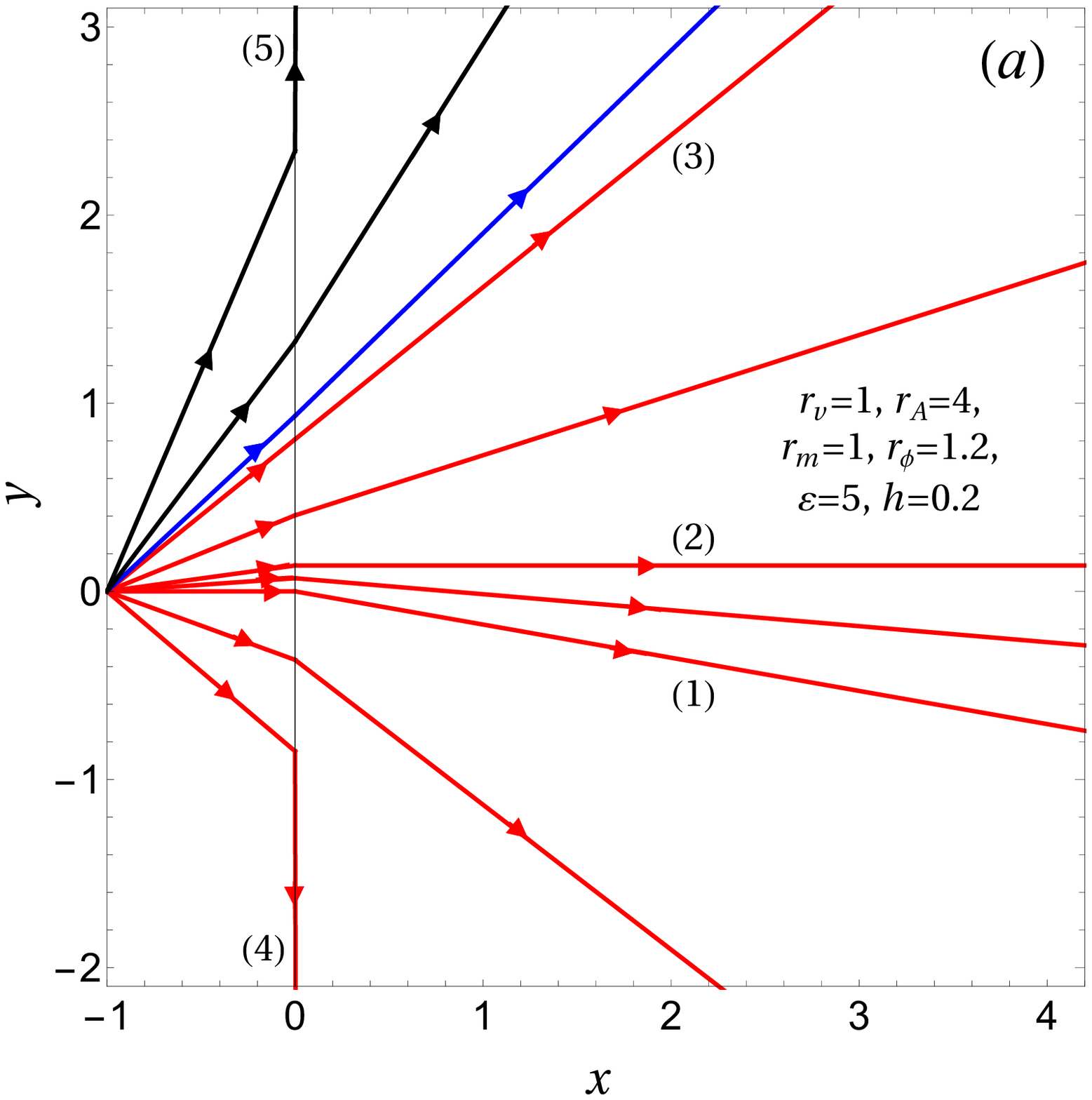}
	\end{minipage}
	\hspace{0.5cm}
	\begin{minipage}[c]{0.3 \textwidth}
		\includegraphics[width=\linewidth]{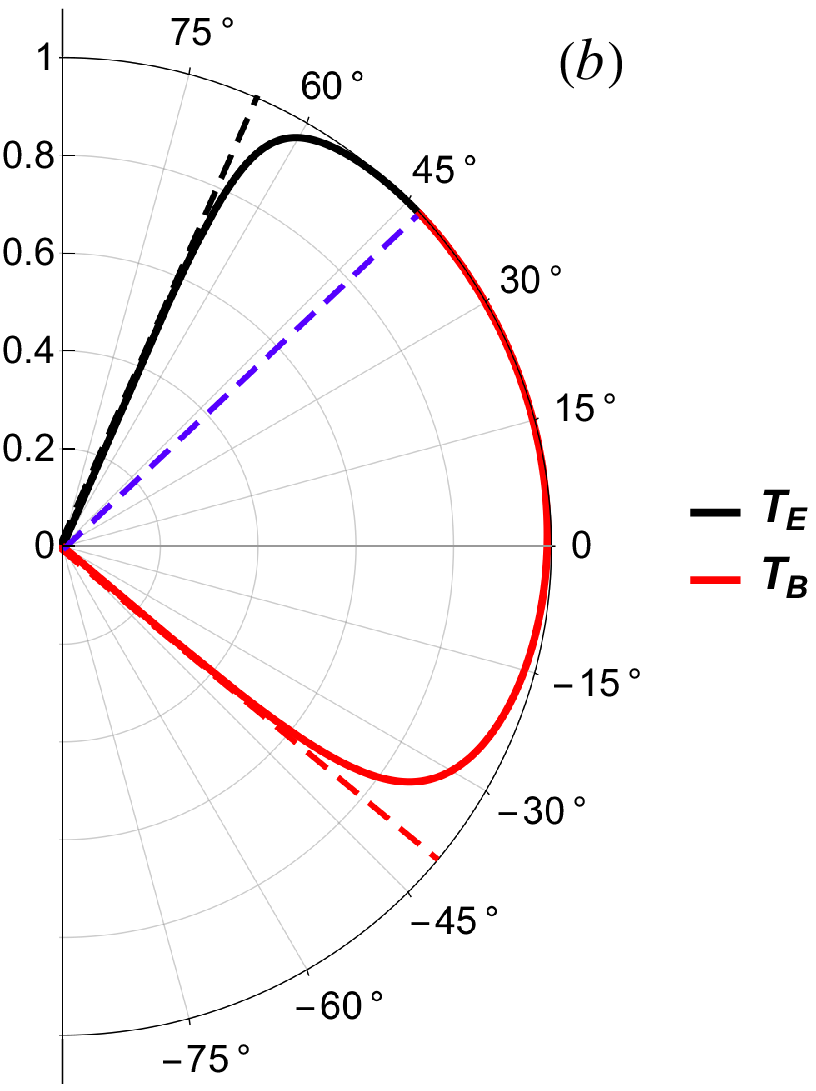}
	\end{minipage}
	\caption{\label{rays}(Color online)(a) The illustration of some rays incident from the left at various incident angles $\alpha$: ray (1) with $\alpha=0^{o}$, $\beta \approx -10 ^{o}$; ray (2) with $\alpha \approx 8 ^{o}$, $\beta=0^{o}$; ray (3) with $\alpha = \beta \approx 39^{o}$; ray (4) with $\alpha = \alpha_{c,-} \approx -40^{o}$, $\beta=-90^{o}$; ray (5) with $\alpha=\alpha_{c,+} \approx 67^{o}$, $\beta=90^{o}$. (b) The plot of transmission probability $T$ as a function of $\alpha$ along with some dashed marking lines. In these figures, the red for $B$-case while the black for the $E$-case, and the blue for the boundary between them.}	
\end{center}
\end{figure}

We can see that except for the vector potential, the other quantities do not qualitatively change the conventional pattern of the refraction. Indeed, it is the term $n_A \equiv h r_A /(1+h)$, originating from the vector potential, that gives rise to the asymmetry of the system concerning the incident angle $\alpha$, which in turn leads to some interesting properties. First, the normal incident $\alpha=0$ corresponds $\beta = \arcsin (-n_A/|\vec{Q}|)$, which is generally non-zero (ray (1) in Figure \ref{rays}a). In contrast, for $\alpha = \arcsin( n_A/|\vec{P}| )$, we have $\beta=0$ (ray (2) in Figure \ref{rays}a). Notably, for any ray between the rays (1) and (2), its incident angle $\alpha$ and the refracted angle $\beta$ are always opposite in sign, which mimics the effect of the negative index of refraction. Note that negative refraction index means our system is probably a Veselago lens when using two beams of Dirac electrons. Second, the ray which goes straight forward without refraction (ray (3) in Figure \ref{rays}a) is no longer the normal ray, instead it corresponds to $\alpha = \beta = \arcsin [ n_A / (|\vec{P}| - |\vec{Q}|)]$. Besides, we also draw the blue ray as the mark for the transition between $B$-case (red rays) and $E$-case (black rays).
Finally, we can define the critical values $\alpha_c$ of the incident angle\footnote{Of course, these definitions only make sense  when $|\sin \alpha_{c,\pm}| \le 1$.} (rays (4) and (5) in Figure \ref{rays}a)
\begin{eqnarray}
	\sin \alpha_{c,\pm} = \left( \pm |\vec{Q}| + n_A \right) / |\vec{P}|.
\end{eqnarray}
 
For incident angles $\alpha$ which are out of the range $(\alpha_{c,-},\alpha_{c,+})$, $q_2$ becomes purely imaginary and the wave-function to the right decays exponentially, making the refracted angle $\beta$ not well-defined any more. Thus the critical values $\alpha_{c,\pm}$ mark the total reflection, or the sharp cut-offs of the transmission probability $T$ when plotted in terms of the incident angle. The plot of $T$ as a function of $\alpha$ is also exhibited in Figure \ref{rays}b, where the critical incident angles are marked by the red (for $\alpha_{c,-}$) and the black (for $\alpha_{c,+}$) dashed lines. Again, we marked the boundary between the $B$- and $E$-cases by the blue line. 

\section{\label{sec:4} Multi-barrier profiles and SUSY transfer matrix method}

\begin{figure}[H]
	\centering
	\includegraphics[width=0.65\linewidth]{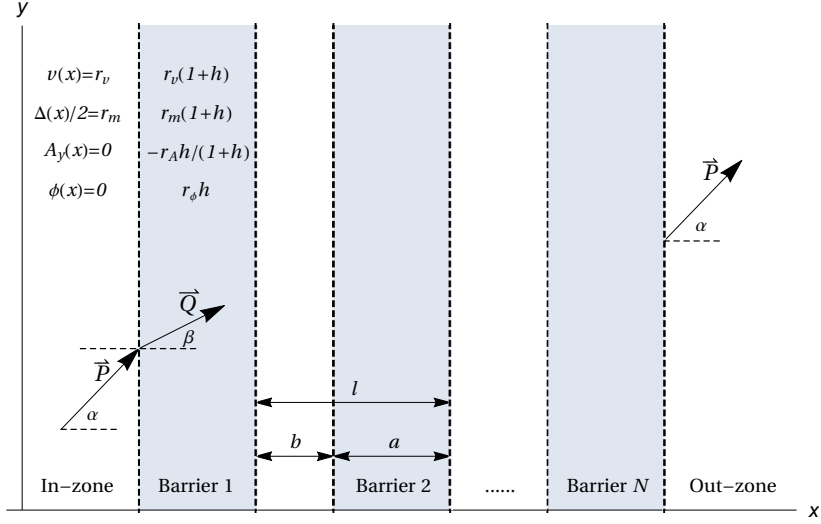}
	\caption{\label{setup-multi}(Color online) Illustration of the experimental setup for the multi-barrier profiles. Each zone is distinguished with the adjacent ones by the thick, dashed, vertical lines. The barriers are shaded. The values of Fermi velocity, band gap, vector and scalar potential in each zone are also given. The beam of charged quasi-particles is assumed to move from the left in the in-zone, go through all the barriers and wells before going out in the out-zone.}
\end{figure}

We are now at the position to develop our formulation for a more complex system, namely a multi-barrier setup as illustrated in Figure \ref{setup-multi}.
Such a configuration can be parameterized as follows:
\begin{eqnarray}
v(x) = r_v \left[ 1 + h \sum_{n=0}^{N-1} \Pi \left(\dfrac{x-n l}{a}\right) \right],~
A_y(x) = r_A / \left[ 1 + h \sum_{n=0}^{N-1} \Pi \left(\dfrac{x-nl}{a}\right) \right] - r_A,\nonumber\\
\Delta(x)/2 = r_m \left[ 1 + h \sum_{n=0}^{N-1} \Pi \left(\dfrac{x-nl}{a}\right) \right],~
\phi(x) = r_\phi \left[ 1 + h \sum_{n=0}^{N-1} \Pi \left(\dfrac{x-nl}{a}\right) \right] - r_\phi,
\end{eqnarray}
where $\Pi(x)$ is the Heaviside Pi function, $N$ is the number of barriers, $l=a+b$ is the spatial period with $a$, $b$ are the widths of the barrier and the well, respectively.
For the sake of convenience, we formally divide the material sheet into $2N+2$ zones which are denoted as below:
\begin{itemize}
	\item The in-zone is labeled by $j=1$;
	\item The barriers are assigned to $j=2,4,6,\dots,2N$;
	\item The wells with $j=3,5,7,\dots,2N+1$ (the last well lies at the right side of the last barrier);
	\item The out-zone with $j=2N+2$.
\end{itemize}
We then can follow the same procedure as in Section \ref{sec:3}. Nevertheless, we now will pay our attention to the transmission probability through the system of barriers and wells rather than the quasi-particle's pseudo-spinor itself. To do that, a slightly modified version of the transfer matrix method (TMM) proposed in \cite{Wang2010} will be utilized to deal with supersymmetric systems. This modified  transfer matrix method, called as Supersymmetric Transfer matrix method or SUSY TMM for short, will be presented in next Subsection. 

\subsection{Supersymmetry Transfer matrix method (SUSY TMM)}
First, we start with $B$-case and recall the necessary results of Section \ref{sec:3} (see \ref{ex2:x->w} for some more details).
Within $j^{th}$ zone, the intermediate pseudo-spinor $G_B$ at two different positions can be rewritten in the following way
\begin{eqnarray}
	G_B (w) &=&
	\begin{pmatrix}
	1 & 1\\
	-i e^{i \theta_j} & i e^{-i \theta_j}
	\end{pmatrix}
	\begin{pmatrix}
	c_j e^{i q_j w} \\
	d_j e^{-i q_j w}
	\end{pmatrix},
	\nonumber\\
	G_B (w+\Delta w) &=&
	\begin{pmatrix}
	e^{i q_j \Delta w} & e^{-i q_j \Delta w} \\
	-i e^{i (\theta_j + q_j \Delta w)} & i e^{-i (\theta_j + q_j \Delta w)}
	\end{pmatrix}
	\begin{pmatrix}
	c_j e^{i q_j w} \\
	d_j e^{-i q_j w}
	\end{pmatrix}.
\end{eqnarray}
where $\theta_j=\theta_2^B$ and $q_j=q_2$ if zone $j$ is a barrier, otherwise $\theta_j=\theta_1^B$ and $q_j=q_1$. Note that in terms of the auxiliary variable $w$, the effective spatial period of the system is $L=A+B$ where $A=a r_v^{-1} (1+h)^{-1}$ and $B=b r_v^{-1}$ are the effective width of barrier and well, respectively. Thus we obtain the relation
\begin{eqnarray}
	G_B (w+\Delta w) 
	= M_j(\Delta w) G_B (w) =
	\dfrac{1}{\cos \theta_j}
	\begin{pmatrix}
	\cos (\theta_j - q_j \Delta w) & - \sin (q_j \Delta w) \\
	\sin (q_j \Delta w) & \cos (\theta_j + q_j \Delta w)
	\end{pmatrix}
	 G_B (w).
\end{eqnarray}
We then can define the transfer matrix $X$ which connects $G_B(w=0)$ (the left edge of the first barrier) and $G_B(w=NL)$ (the right edge of the last well):
\begin{eqnarray}\label{X}
	G_B (N L)
	= \left[ \prod_{j=2N+1}^{2} M_j (z_j) \right] G_B(0)
	= X G_B(0),
\end{eqnarray}
where the width of zone $j$ is $z_j=A$ if $j$ is even, or $z_j=B$ if $j$ is odd.
Because all the $N$ barriers as well as all the $N$ wells are the same, we can simplify the expression of $X$ matrix as
\begin{eqnarray}
	X = \left[ M_3 (B) M_2 (A) \right]^N.
\end{eqnarray}

The matching conditions at all the boundaries, which has been implied in Equation \eqref{X} already, can be written explicitly as
\begin{eqnarray}
	\begin{pmatrix}
	c_{2N+2} e^{i q_1 N L} \\
	- i e^{i \theta_1^B} c_{2N+2} e^{i q_1 N L}
	\end{pmatrix}
	=
	\begin{pmatrix}
	X_{11} & X_{12} \\
	X_{21} & X_{22}
	\end{pmatrix}
	\begin{pmatrix}
	c_1 + d_1 \\
	- i e^{i \theta_1^B} c_1 + i e^{-i \theta_1^B} d_1
	\end{pmatrix}.
\end{eqnarray}
Solving this gives us the expression of the transmission coefficient in this case
\begin{eqnarray}
	t_B \equiv \dfrac{c_{2N+2}}{c_1} = \dfrac{2 \exp \left(-i q_1 N L\right) \cos \theta_1^B}{e^{i \theta_1^B} X_{11} + e^{-i \theta_1^B} X_{22} + i (X_{12} - X_{21})}~.
\end{eqnarray}
Here we made use of the fact that $\det X = 1$ (resulting from the property $\det M_j = 1$) to shorten the above expression.

By the same procedure, we can obtain the transmission coefficient for $E$-case
\begin{eqnarray}\label{t<}
	t_E = \dfrac{2 \exp \left(-i q_1 N L\right) \cos \theta_1^E}{e^{i \theta_1^E} X_{11} + e^{-i \theta_1^E} X_{22} - (X_{12} + X_{21})},
\end{eqnarray}
where the $X$ matrix is now built from
\begin{eqnarray}
M_j (\Delta w) = \dfrac{1}{\cos \theta_j}
\begin{pmatrix}
\cos (\theta_j - q_j \Delta w) & i \sin (q_j \Delta w) \\
i \sin (q_j \Delta w) & \cos (\theta_j + q_j \Delta w)
\end{pmatrix}.
\end{eqnarray}
To sum up, with a given multi-barrier setup, we can compute the $X$ matrix and then easily obtain the transmission probability $T=|t_{B(E)}|^2$.

\subsection{Profile of the transmission probability}

\begin{figure}[H]
	\centering
	\hfill
	\begin{minipage}[c]{0.3 \textwidth}
		\includegraphics[width=\linewidth]{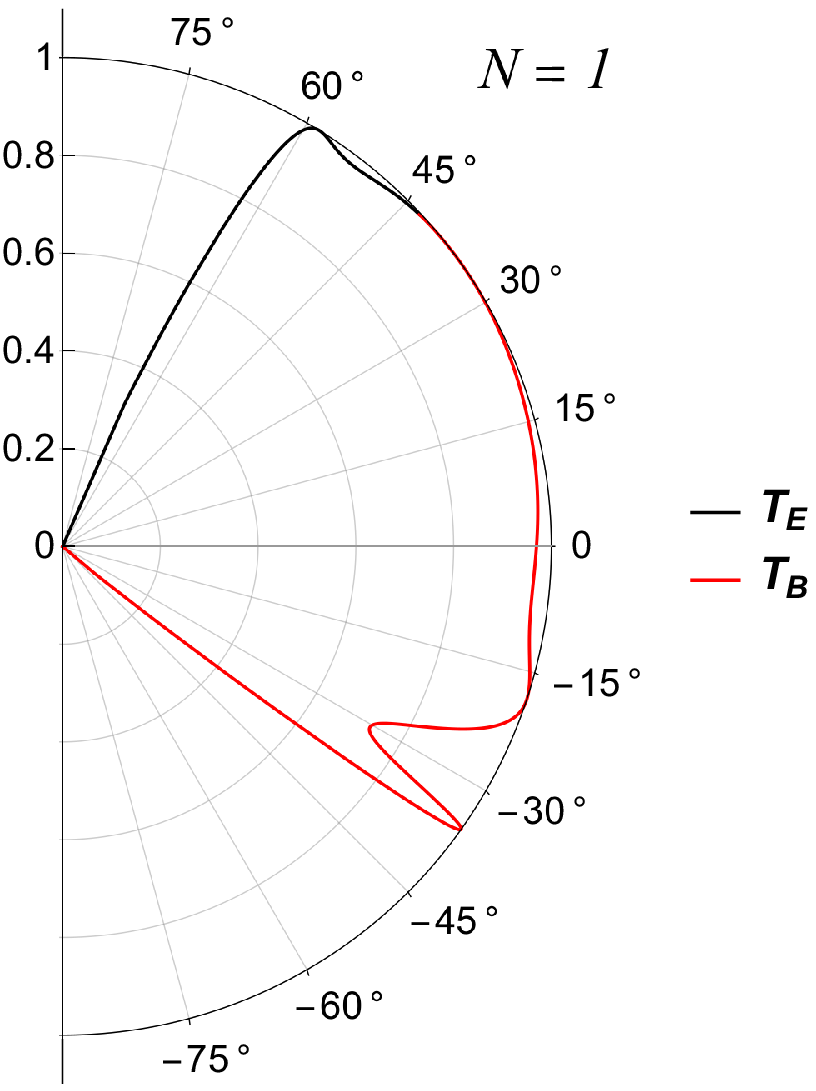}
	\end{minipage}
	\hfill
	\begin{minipage}[c]{0.3 \textwidth}
		\includegraphics[width=\linewidth]{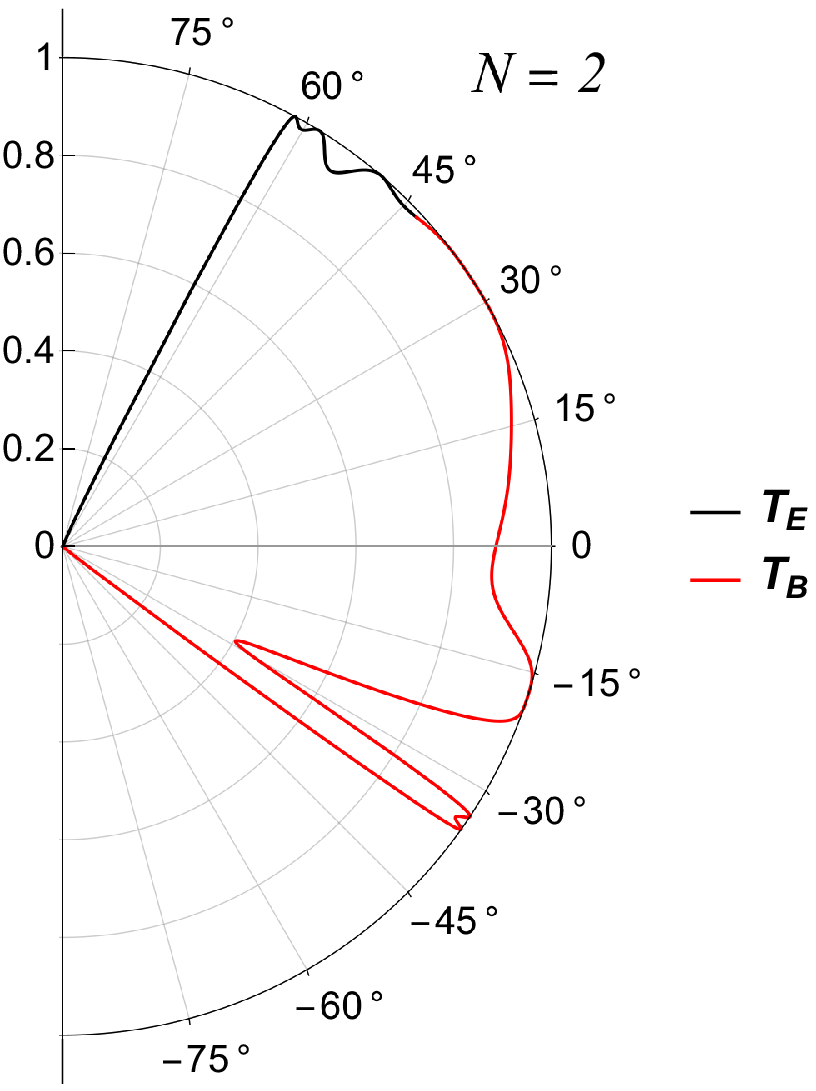}
	\end{minipage}
	\hfill
	\begin{minipage}[c]{0.3 \textwidth}
		\includegraphics[width=\linewidth]{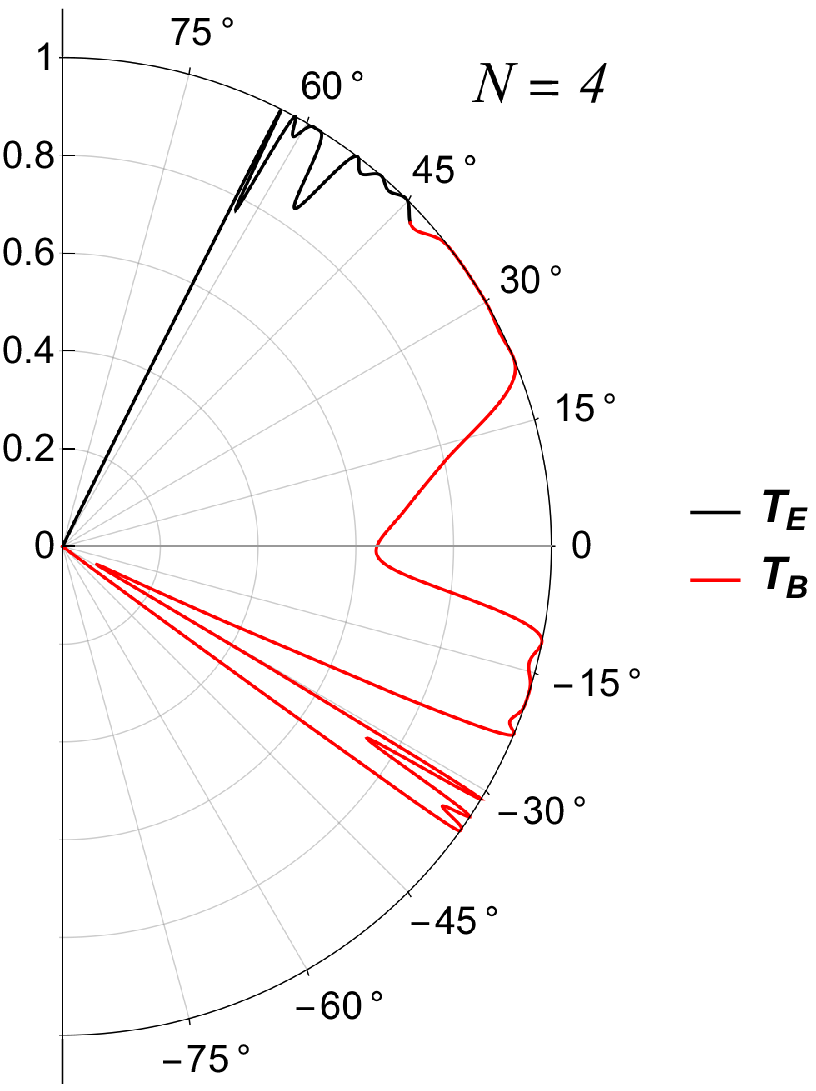}
	\end{minipage}
	\\
	\hfill
	\begin{minipage}[c]{0.3 \textwidth}
		\includegraphics[width=\linewidth]{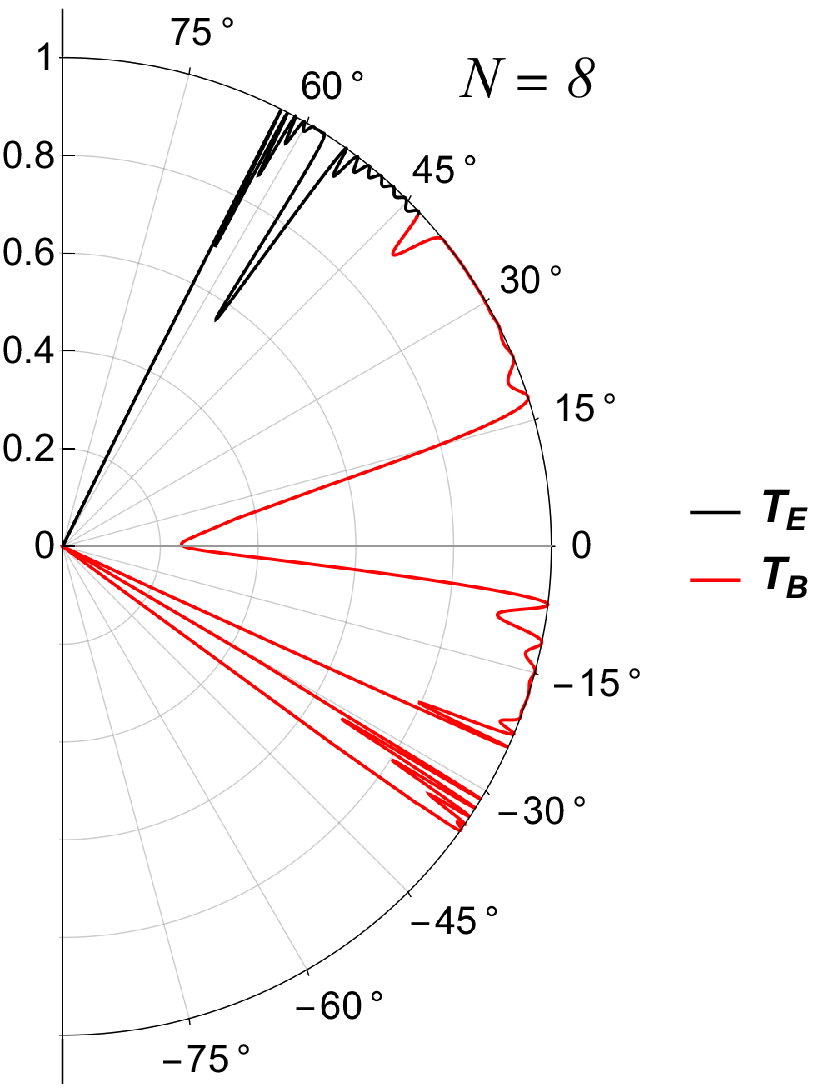}
	\end{minipage}
	\hfill
	\begin{minipage}[c]{0.3 \textwidth}
		\includegraphics[width=\linewidth]{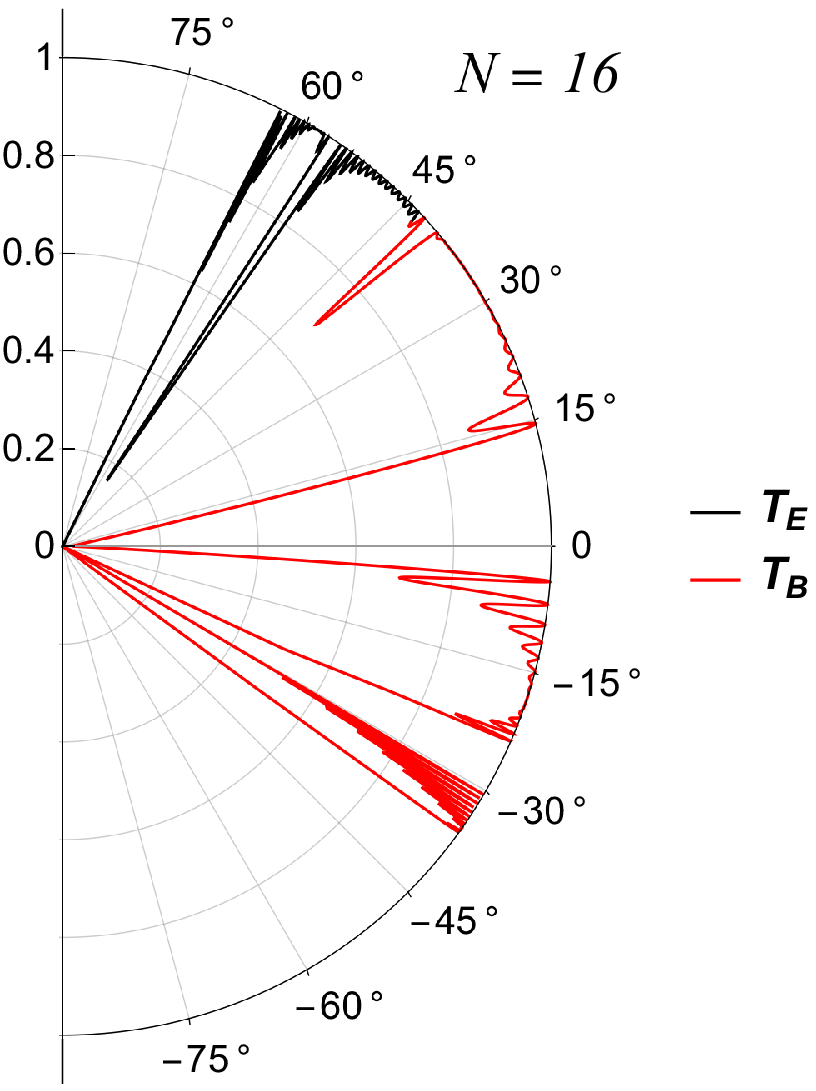}
	\end{minipage}
	\hfill
	\begin{minipage}[c]{0.3 \textwidth}
		\includegraphics[width=\linewidth]{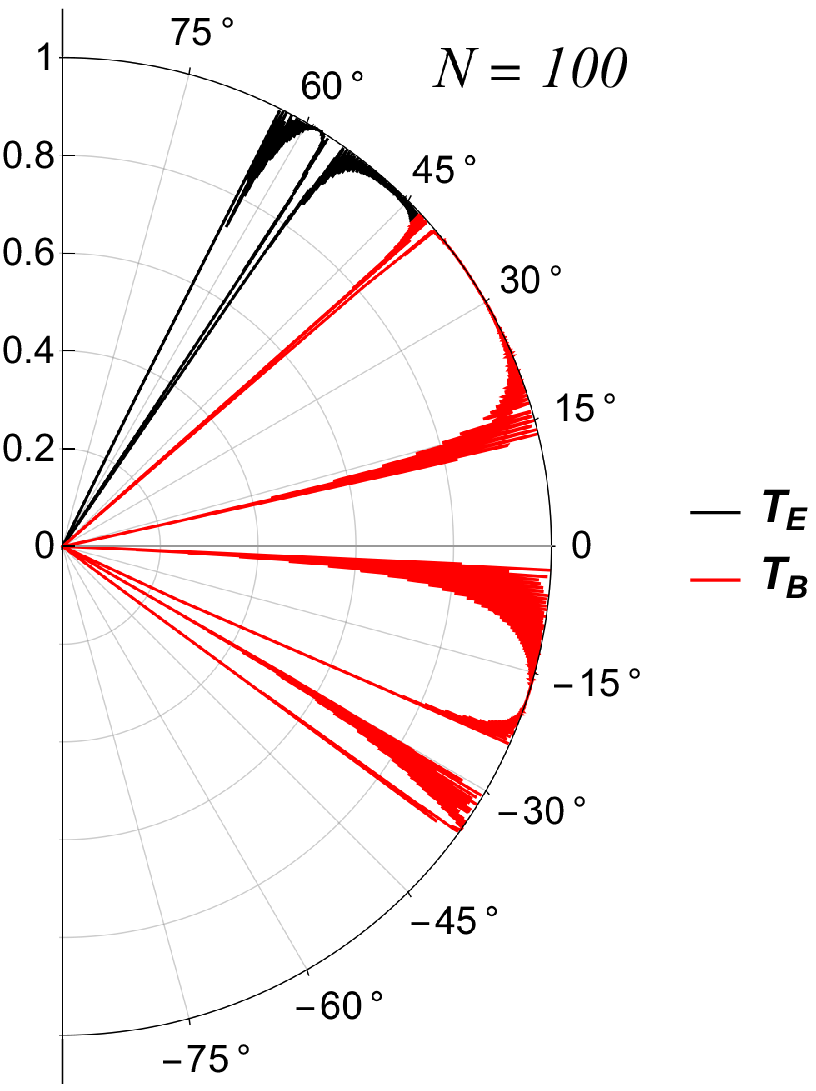}
	\end{minipage}
	\caption{\label{trans-multi}(Color online) Plots of the transmission probability $T$ as a function of $\alpha$ for different numbers $N$ of barriers. Except the new input parameters $b=1$ and $N=1,2,4,8,16,100$, the others are the same as in Figure \ref{rays}. It is exhibited that the more barriers, the more quickly $T(\alpha)$ varies and the more transparent peaks exist. Eventually, when $N$ is very large, the transparent-bands of $\alpha$ form, interposed by the gap-bands. Note that the cut-offs $\alpha_{c,-} \approx -40^\circ$ and $\alpha_{c,+} \approx 67^\circ$ are the same as those in Figure \ref{rays} of the step profiles, no matter the number $N$ of barriers.}
\end{figure}

Figure \ref{trans-multi} shows the transmission probability through the multi-barrier configuration as a function of the incident angle $\alpha$ with some different numbers $N=1,2,4,8,16,100$ of the barriers. We see that as $N$ increases, there are more transparent peaks where $T=1$ (this can be explained by some mathematical manipulations, see \ref{peaks} for details) while some of the minima reach closely to zero. Consequently, when $N$ is large enough, we will obtain some transparent-bands (domains of $\alpha$ at which the quasi-particle can transmit completely through the system), distinguished to each other by the angular gap-bands where the transmission of the quasi-particle is almost banned. Note that the plots in Figure \ref{trans-multi} use the same parameters as in Figure \ref{rays} and we can observe the same cut-offs $\alpha_{c,-} \approx -40^\circ$ and $\alpha_{c,+} \approx 67^\circ$. So the number $N$ of barriers does nothing with these cut-offs, apart from making them sharper, as expected. This means to manipulate the angular positions of these cut-offs, we have to tune the other system's parameters instead. Also because of the optic-like behavior of Dirac electron in the situation of single barrier $N = 1$, the large-$N$ system could be a Veselago lens or Fabry–-P{\'e}rot cavity (see Reference \cite{Wilmart2014} and references therein for examples).

\section{\label{sec:5}Discussion on Klein tunnelling}

Finally, we examine the Klein tunnelling in multi-barrier configuration. More particularly, we focus on whether the transmission probability is unity when the incident wave comes at the right angle $\alpha=0$.

First, we consider the simplest situation of massless Dirac fermion and no vector potential, i.e. $r_m=-q_m=0$ and $r_A=-q_A=0$. Then because $|q_\phi|>|q_m|$, a Dirac fermion incident with $\alpha=0$ behaves as in $E$-case. Besides, both $q_1 = |\varepsilon|$ and $q_2 = |\varepsilon - r_\phi h|$ are real while it follows from Equation \eqref{q1q2-ele} that $\theta_1^E=\theta_2^E=0$. All of these simplify the expression of the transmission coefficient at the right angle:
\begin{eqnarray}\label{Klein}
	t_{E}(\alpha=0) = e^{i (q_2 - q_1) N w_0}.
\end{eqnarray}
So when the quasi-particle is massless and does not experience a vector potential, the transmission probability at the right angle is always unity, $T(\alpha=0)=|t_{E}(\alpha=0)|^2=1$, regardless how high or how long the scalar potential- and velocity-barriers are as well as the number of barriers. This is a manifestation of the Klein tunnelling \cite{Calogeracos1999}. 

\begin{figure}[H]
	\centering
	\hfill
	\begin{minipage}[c]{0.45 \textwidth}
		\includegraphics[width=\linewidth]{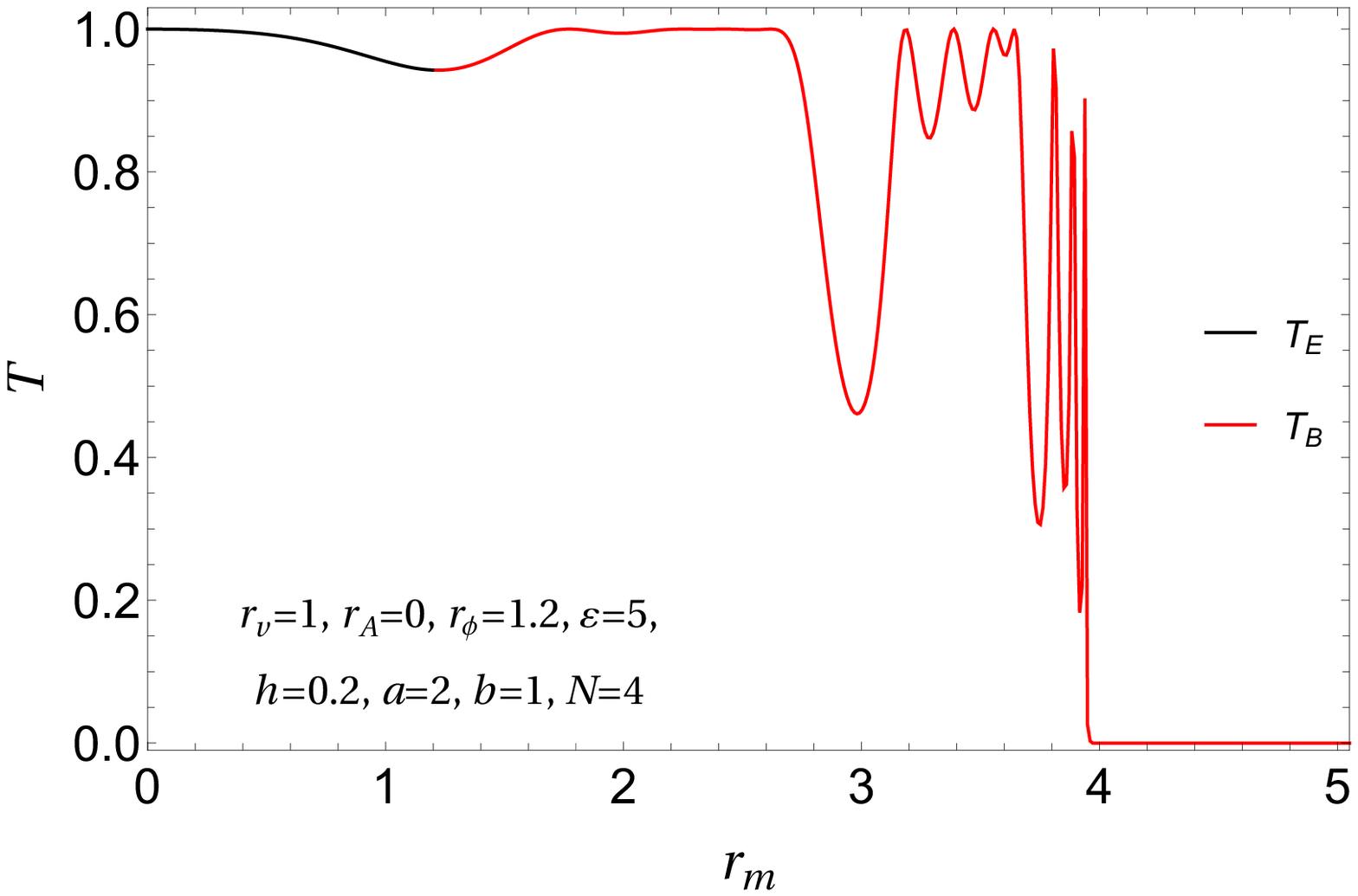}
	\end{minipage}
	\hfill
	\begin{minipage}[c]{0.45 \textwidth}
		\includegraphics[width=\linewidth]{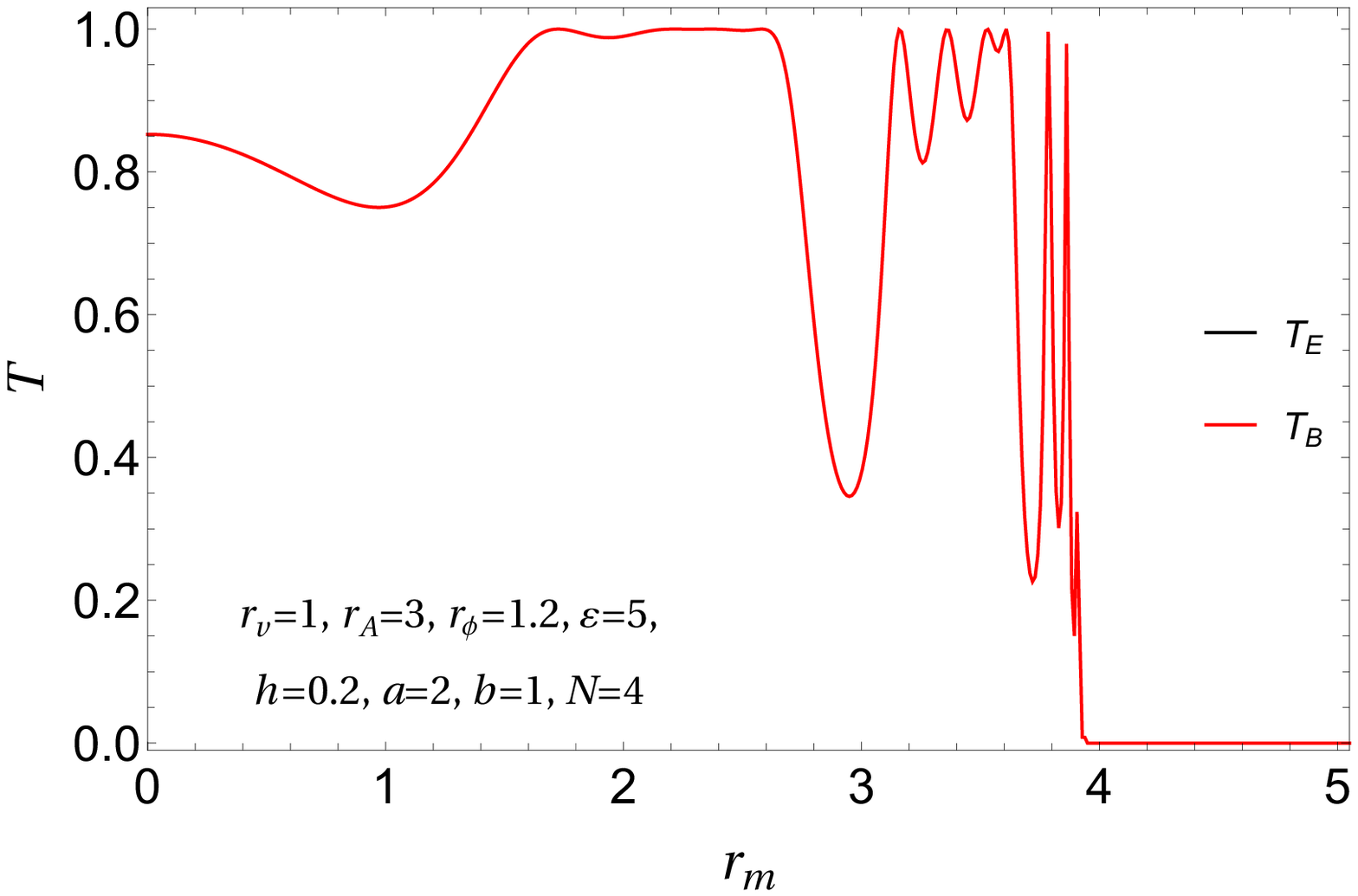}
	\end{minipage}
	\caption{\label{T(rm)}(Color online) Plots of the transmission probability as a function of $r_m$ while $r_A=0$ (left) or $r_A \ne 0$ (right). Other parameters are: $r_v=1$, $r_\phi=1.2$, $\varepsilon=5$, $h=0.2$, $a=2$, $b=1$, $N=4$.}
\end{figure}

\begin{figure}[H]
	\centering
	\hfill
	\begin{minipage}[c]{0.45 \textwidth}
		\includegraphics[width=\linewidth]{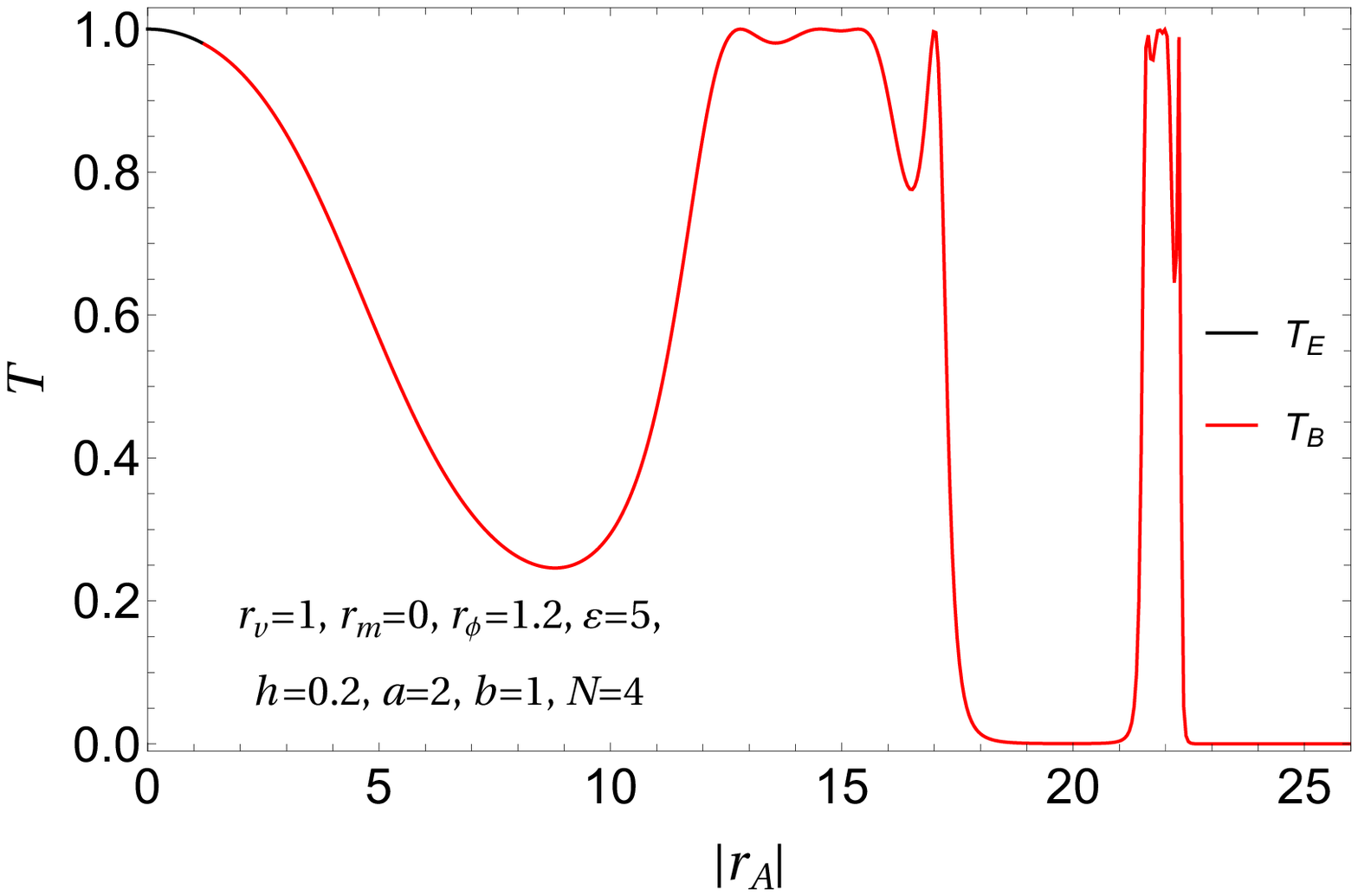}
	\end{minipage}
	\hfill
	\begin{minipage}[c]{0.45 \textwidth}
		\includegraphics[width=\linewidth]{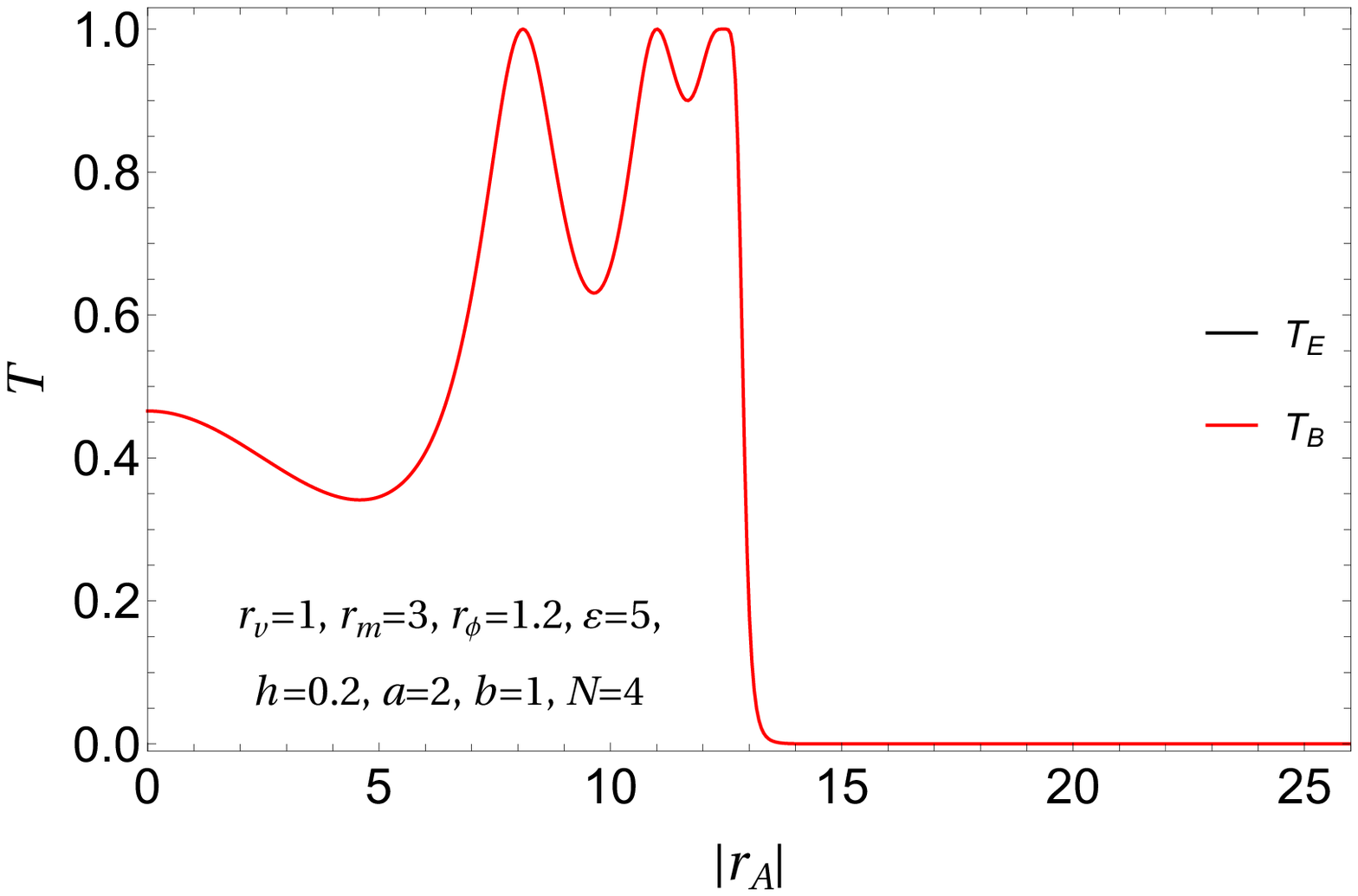}
	\end{minipage}
	\caption{\label{T(ra)}(Color online) Plots of the transmission probability as a function of $r_A$ while $r_m=0$ (left) or $r_m \ne 0$ (right). Other parameters are: $r_v=1$, $r_\phi=1.2$, $\varepsilon=5$, $h=0.2$, $a=2$, $b=1$, $N=4$.}
\end{figure}

Next, we move to the situation in which either $r_m$ or $r_A$ or both of them do(es) not vanish. To analyze this situation, we calculate the transmission probability $T$ when changing the value of $r_m$ (see Figure \ref{T(rm)}) or $r_A$ (see Figure \ref{T(ra)}) and keeping the others fixed.

As can been seen from the figures, we have $T=1$ when $r_m=r_A=0$ as expected. Notably, the transmission probability can also be unity with many other pairs of values $(r_m,r_A)$. In comparison with the conclusion in \cite{Peres2009}, in which the authors claimed that the effective mass and the vector potential can prevent the Klein tunnelling while the Fermi velocity and the scalar potential can not, our results in this work showed a bit more general conclusion that the Klein tunnelling can resurge even when either the effective rest mass of the quasi-particle or the vector potential or both of them present(s). Moreover, we can again identify the cut-offs, points at which the transmission probability suddenly drops to almost zero because the wave function of the quasi-particle becomes exponentially decaying upon transporting through the barriers. These cut-offs may be useful for controlling, i.e turning on and off, the currents of charged quasi-particles through the multi-barrier system, which is promising for electronic applications.

\section{\label{sec:6}Conclusion}

In this work, a  system of a $2D$ Dirac material sheet with step and multi-barrier profiles of Fermi velocity, effective mass, magnetic and electric fields has been analysed. Such a system can be experimentally produced using an electric cavity, a series of in-plane electrodes, superconducting materials combined with an appropriate strain engineering. We used the supersymmetric formalism, in which we took a suitable ansatz into account, to obtain the analytical expression of the wave function in terms of the quasi-particle's energy in the case of step profiles. Then, we developed the method further by integrating it with the transfer matrix method (that we called SUSY TMM). This approach proved its usefulness for  multi-barrier profiles, as it gave us a straightforward way to calculate the transmission probability of the quasi-particle transporting through the system. Considering the simple situation of step profiles allowed us to deduce a Snell-like relation for the transportation of the charged quasi-particle between two different zones of the sheet. The observation of optic-like behaviour of quasi-particle in our system suggests that our system could be used as Veselago lens. Meanwhile the multi-barrier system may be a system of Veselago lens or a Fabry-–P{\'e}rot cavity. Then the calculations for multi-barrier profiles revealed the tendency of the transmission probability when the number of barriers increases: when there are a lot of barriers, we will achieve a band-like pattern. This pattern consists of angular domains in which the transmission is highly perfect ($T > 0.9$), separated by angular gaps in which the transmission is mostly banned ($T < 0.1$). Also, we successfully reproduced the Klein tunnelling and reclaimed the uselessness of the Fermi velocity and scalar potential barriers in suppressing this effect. Further, we showed that even with the presence of the effective mass and the vector potential, the Klein tunnelling may still exist. Last but not least, our examination pointed out that the cut-offs in the angular distribution of the transmission probability can be manipulated by some of the parameters of the system, suggesting the potential applications in electronics. We believe that our results exhibited in this work are useful in both methodological as well as practical aspects. In the spirit of References \cite{LE2019PhysicaE, Le2019JPCM, Phan2019, Le2020JPCM, Fernandez2020, Wagner2020, Ocampo2021}, the similar framework may be developed for other geometrical forms of $2D$ Dirac materials such as fullerene, carbon nanotube, carbon nanohorn, carbon pseudosphere, etc. or bilayer, multilayer graphene or even organic Dirac materials with titled Dirac cone.

\section*{Acknowledgement}
One of the authors, Dai-Nam Le, was funded by Vingroup Joint Stock Company and supported by the Domestic Master/ PhD Scholarship Programme of Vingroup Innovation Foundation (VINIF), Vingroup Big Data Institute (VINBIGDATA), code: VINIF.2020.TS.03. The author Anh-Luan Phan would like to express his great gratitude to his beloved mother for her support in the period of time he conducted this work. The authors also thank Professor Van-Hoang Le (Department of Physics, Ho Chi Minh City University of Education, Vietnam) for encouragement and Professor Pinaki Roy (Atomic Molecular and Optical Physics Research Group, Advanced Institute of Materials Science, Ton Duc Thang University, Ho~Chi~Minh~City, Vietnam) for suggesting the problem and going through the manuscript.

\section*{Author contribution statement}
All authors contributed equally to the paper. All the authors have read and approved the final manuscript.

\bibliographystyle{elsarticle-num}
\bibliography{scat2020}

\pagebreak
\appendix

\section{\label{rotation} The pseudo-spinor rotations}

A general rotation $U(\alpha,\beta,\gamma)$ of the pseudo-spinor has the form
\begin{eqnarray}
	K(w) = U G(w) =
	\exp \left[\alpha/2 \left( \sigma_y \cos \beta + \sigma_z \sin \beta \cos \gamma + \sigma_x \sin \beta \sin \gamma \right) \right]
	G(w),
\end{eqnarray}
where $\alpha$, $\beta$ and $\gamma$ are the three degrees of freedom (the fourth one is canceled by imposing the constraint $\det U = 1$). Using this rotation, we can transform the equation for $K(w)$ into an equation for $G(w)$
\begin{eqnarray}
\left\{-i \sigma_x \partial_w + \sigma_y \left[ \tilde{k} + \tilde{A}(w) \right] + \sigma_z \tilde{\Delta}(w)/2 + \tilde{\phi}(w) - \tilde{\varepsilon} \right\} G(w) = 0,
\end{eqnarray}
with
\begin{eqnarray}
&&\tilde{A}(w) = \bigg[\sinh^2 \dfrac{\alpha}{2} \sin^2\beta \left(i q_\phi \sin 2\gamma - q_v (k+A_0) \cos 2\gamma \right) \nonumber\\
&&\qquad\qquad + \sinh^2 \dfrac{\alpha}{2} \left( \dfrac{1}{2} q_v (k+A_0) \cos 2\beta + q_m \sin 2\beta \cos\gamma \right)  \nonumber\\
&&\qquad\qquad + \sinh\alpha (q_\phi \cos\beta + i q_m \sin\beta \sin\gamma) + \dfrac{1}{4} q_v (k+A_0) (1 + 3 \cosh\alpha ) \bigg] p(w), \nonumber\\
&&\tilde{\phi}(w) = \bigg[ \sinh^2 \dfrac{\alpha}{2} \sin ^2\beta ( q_\phi \cos 2\gamma - i q_v (k+A_0) \sin 2\gamma ) \nonumber\\
&&\qquad\qquad + \sinh\alpha (q_v (k+A_0) \cos\beta +q_m \sin\beta  \cos\gamma) \nonumber\\
&&\qquad\qquad + \sinh^2 \dfrac{\alpha}{2} \left( \dfrac{1}{2} q_\phi \cos 2\beta + i q_m \sin 2\beta  \sin\gamma \right) + \dfrac{1}{4} q_\phi (1 + 3 \cosh\alpha ) \bigg] p(w), \nonumber\\
&&\dfrac{\tilde{\Delta}(w)}{2} = \bigg[\sinh^2 \dfrac{\alpha}{2}  \left(q_v (k+A_0) \sin 2\beta \cos\gamma -i q_\phi \sin 2\beta \sin\gamma -q_m \cos 2\beta \right) \nonumber\\
&&\qquad\qquad + \sinh\alpha  \sin\beta  \left(q_\phi \cos\gamma -i q_v (k+A_0) \sin\gamma \right) + \cosh^2 \dfrac{\alpha}{2} q_m\bigg] p(w) \nonumber\\
&&\qquad\qquad + \sinh\alpha \sin\beta \left((\phi_0 - \varepsilon) \cos\gamma - i q_v q_A \sin\gamma \right) \nonumber\\
&&\qquad\qquad + \sinh^2 \dfrac{\alpha}{2} \sin 2\beta (q_v q_A \cos\gamma - i (\phi_0 - \varepsilon) \sin\gamma) , \nonumber\\
&&\tilde{k} = + \sinh^2 \dfrac{\alpha}{2} \left[\dfrac{1}{2} q_v q_A \cos 2\beta - \sin ^2\beta (q_v q_A \cos 2\gamma -i (\phi_0 - \varepsilon) \sin 2\gamma )\right] \nonumber\\
&&\qquad\qquad + (\phi_0 - \varepsilon) \sinh\alpha  \cos\beta + \dfrac{1}{4} q_v q_A (1 + 3\cosh\alpha) , \nonumber\\
&&\tilde{\varepsilon} = - \sinh^2 \dfrac{\alpha}{2} \left[ \dfrac{1}{2} (\phi_0 - \varepsilon) \cos 2\beta + \sin ^2\beta  \left((\phi_0 - \varepsilon) \cos 2\gamma -i q_v q_A \sin 2\gamma \right)\right] \nonumber\\
&&\qquad\qquad- q_v q_A \sinh\alpha  \cos\beta - \dfrac{1}{4} (\phi_0 - \varepsilon) (1 + 3\cosh\alpha).
\end{eqnarray}

To obtain the $B$-case, we can use the rotation $U_B (\alpha^B,\beta^B,\gamma^B)$ with
\begin{eqnarray}
	\cosh \dfrac{\alpha^B}{2} &=& \dfrac{s_1^B \left[ (k+A_0)q_v + \sqrt{-Q} \right] - i s_2^B (q_m+q_\phi)}{2}, \qquad\qquad\qquad\quad\qquad~~ \nonumber\\
	\cos\beta^B &=& i\dfrac{s_2^B \left[ (k+A_0)q_v - \sqrt{-Q} \right] + i s_1^B (q_m+q_\phi)}{2 \sinh \dfrac{\alpha^B}{2}}, \nonumber\\
	\cos\gamma^B &=& \dfrac{s_1^B \left[ (k+A_0)q_v + \sqrt{-Q} \right] + i s_2^B (q_m+q_\phi)}{2 \sin\beta^B \sinh \dfrac{\alpha^B}{2}}.
\end{eqnarray}
For $E$-case, the rotation $U_E$ is now determined by
\begin{eqnarray}
	\cosh \dfrac{\alpha^E}{2} &=& \dfrac{i (k+A_0)q_v (s_1^E+s_2^E) - \sqrt{Q}(s_1^E-s_2^E) + (q_m+q_\phi)(s_1^E-s_2^E)}{2}, \nonumber\\
	\cos \beta^E &=& i \dfrac{i (k+A_0)q_v (s_1^E-s_2^E) - \sqrt{Q}(s_1^E+s_2^E) - (q_m+q_\phi)(s_1^E+s_2^E)}{2 \sinh \dfrac{\alpha^E}{2}}, \nonumber\\
	\cos \gamma^E &=& \dfrac{i (k+A_0)q_v (s_1^E+s_2^E) - \sqrt{Q}(s_1^E-s_2^E) - (q_m+q_\phi)(s_1^E-s_2^E)}{2 \sin\beta^E \sinh \dfrac{\alpha^E}{2}}.
\end{eqnarray}
Here, the factors $s$ are given by\footnote{The signs of $s_{1,2}^{B,E}$ must satisfy the constraint $\det U=1$, or equivalently, $(s_1^B s_2^B)^{-1} = -2 (q_m+q_\phi) \sqrt{Q}$ for $B$-case and $(s_1^E s_2^E)^{-1} = 4 (q_m+q_\phi) \sqrt{Q}$ for $E$-case.}
\begin{eqnarray}
(s_{1,2}^{B})^2 = \dfrac{ \sqrt{\pm 1} }{ \sqrt{\mp 1} } \left[ 2 (q_m+q_\phi) \sqrt{Q} \right]^{-1} \times \qquad\qquad\qquad\qquad\qquad\qquad\qquad\qquad\qquad\qquad\qquad\qquad \nonumber\\
	\dfrac{  \sqrt{q_v \left[(k+A_0) q_v \mp \sqrt{-Q}\right] \left[(\varepsilon-\phi_0) (k+A_0) + q_A (q_m+q_\phi)\right] + (\varepsilon-\phi_0) q_m (q_m + q_\phi)} }{ \sqrt{q_v \left[(k+A_0) q_v \pm \sqrt{-Q}\right] \left[(\varepsilon-\phi_0) (k+A_0) + q_A (q_m+q_\phi)\right] + (\varepsilon-\phi_0) q_m (q_m + q_\phi)} },\quad \nonumber\\
(s_{1,2}^{E})^2 = \dfrac{ \sqrt{\pm 1} }{ \sqrt{\mp 1} } \left[ 4 (q_m+q_\phi) \sqrt{Q} \right]^{-1} \times \qquad\qquad\qquad\qquad\qquad\qquad\qquad\qquad\qquad\qquad\qquad\qquad \nonumber\\
	\dfrac{ \sqrt{q_v \left[(k+A_0) q_v \mp i \sqrt{Q} \right] \left[(\varepsilon-\phi_0) (k+A_0) + q_A (q_m + q_\phi)\right] + (\varepsilon-\phi_0) q_m (q_m + q_\phi)} }{ \sqrt{q_v \left[(k+A_0) q_v \pm i \sqrt{Q} \right] \left[(\varepsilon-\phi_0) (k+A_0) + q_A (q_m + q_\phi)\right] + (\varepsilon-\phi_0) q_m (q_m + q_\phi)} }.\quad
\end{eqnarray}

\section{\label{rho&J} The formulae of the probability density and the probability current density}

For both the $B$- and $E$-cases, the expressions of the probability density $\rho$ and the probability current density $\vec{J}$ are given by \cite{Peres2009}
\begin{eqnarray}
\rho^{B,E} (x) &=& \Psi^\dagger (x,y) \Psi(x,y) = \dfrac{1}{v(x)} G_{B,E}^\dagger(w(x)) U_{B,E}^\dagger U_{B,E} G_{B,E}(w(x)),\nonumber\\
J_x^{B,E} &=& v(x) \Psi^\dagger (x,y) \sigma_x \Psi(x,y) = G_{B,E}^\dagger(w(x)) U_{B,E}^\dagger \sigma_x U_{B,E} G_{B,E}(w(x)),\nonumber\\
J_y^{B,E} &=& v(x) \Psi^\dagger (x,y) \sigma_y \Psi(x,y) = G_{B,E}^\dagger(w(x)) U_{B,E}^\dagger \sigma_y U_{B,E} G_{B,E}(w(x)).
\end{eqnarray}

\section{\label{ex2:x->w}Variable-changing for multi-barrier system}
The auxiliary variable is
\begin{eqnarray}
w &=& \int_{x_0 - a/2}^{x} \dfrac{\text{d}u}{r_v \left[ 1 + h \sum_{n=0}^{N-1} \Pi \left(\dfrac{x-nl}{a}\right) \right]} \nonumber\\
&=& \dfrac{1}{r_v} \times
\begin{cases}
x + a/2 \qquad\qquad\qquad\qquad\qquad~ \text{ in zone $j=1$;}\\
\dfrac{x + a/2}{1+h} + (j/2-1) \dfrac{b h}{1+h} ~~~~~~~ \text{ in zone $j=2,4,\dots,2N$;}\\
x + a/2 + a \left( 1 - \dfrac{h (j-1)/2}{1+h} \right) ~\text{ in zone $j=3,5,\dots,2N+1$;}\\
x + a/2 + a \left( 1 - \dfrac{N h}{1+h} \right) \qquad~~ \text{ in zone $j=2N+2$.}
\end{cases}
\end{eqnarray}
We can rewrite the above quantities to show that they satisfy the ansatz \eqref{ansatz} with
\begin{eqnarray}
&&q_v = -r_v < 0, \quad q_A = -r_A, \quad q_m = -r_m \le 0, \quad q_\phi = - r_\phi, \nonumber\\
&&p(w) = - \left[ 1 + h \sum_{n=0}^{N-1} \Pi \left(\dfrac{w - n L}{A} -\dfrac{1}{2} \right)  \right] < 0.
\end{eqnarray}

\section{\label{peaks} The increase in the number of transparent peaks when $N$ increases}

To explain the increase in the number of transparent peaks when the number $N$ of barriers increases, we re-examine the transfer matrix $X$. Because, in principle, all $2\times 2$ matrices can be uniquely decomposed into the Pauli matrices $\sigma_j$ ($j=1,2,3$) and the identity matrix $\mathbb{I}$, we have
\begin{eqnarray}
X = [M_3 M_2]^N &=& [A_0 \mathbb{I} + A_1 \sigma_1 + A_2 \sigma_2 + A_3 \sigma_3]^N\nonumber\\
&=& \left[ u_0 + \cos (u) + i \sin(u) \left( \dfrac{u_1}{u} \sigma_1 + \dfrac{u_2}{u} \sigma_2 + \dfrac{u_3}{u} \sigma_3 \right) \right]^N
\end{eqnarray}
where we introduced the vector $\vec{u}=(u_1,u_2,u_3)$ whose components satisfy the following relations
\begin{eqnarray}
i u_j \sin(u) /u = A_j, \qquad j=1,2,3.
\end{eqnarray}
Here $u=\sqrt{u_1^2+u_2^2+u_3^2}$ is the length of $\vec{u}$ and in this situation $\cos(u) = A_0 - u_0$.
Then, according to Euler's identity for matrix, $X$ can be rewritten
\begin{eqnarray}
X &=& \left[ u_0 + e^{i \vec{u} \cdot \vec{\sigma}} \right]^N \nonumber\\
&=& \sum_{n=0}^{N} u_0^{N-n} e^{i n \vec{u} \cdot \vec{\sigma}} \nonumber\\
&=& \sum_{n=0}^{N} u_0^{N-n} \left[ \cos (n u) + i \sin(n u) \left( \dfrac{u_1}{u} \sigma_1 + \dfrac{u_2}{u} \sigma_2 + \dfrac{u_3}{u} \sigma_3 \right) \right] \nonumber\\
&=& X_0(N) + i \left( \dfrac{u_1}{u} \sigma_1 + \dfrac{u_2}{u} \sigma_2 + \dfrac{u_3}{u} \sigma_3 \right) X_\sigma(N)
\end{eqnarray}
where
\begin{eqnarray}
X_0(N) &=& \sum_{n=0}^{N} u_0^{N-n} \cos (n u) = \frac{\cos (N u) - u_0\cos [(N+1) u] - u_0^{N+1} \cos (u) +u_0^{N+2}}{\left(e^{i u} - u_0\right) \left(e^{-i u} - u_0\right)}, \nonumber\\
X_\sigma(N) &=& \sum_{n=0}^{N} u_0^{N-n} \sin (n u) = \frac{\sin (N u) - u_0 \sin [(N+1)u] + u_0^{N+1} \sin(u)}{\left(e^{i u} - u_0\right) \left(e^{-i u} - u_0\right)}.
\end{eqnarray}
Or in matrix form, we have
\begin{eqnarray}
X =
\begin{pmatrix}
X_0 (N) + i X_\sigma(N) u_3/u & i X_\sigma(N) u_1/u + X_\sigma(N) u_2/u \\
i X_\sigma(N) u_1/u - X_\sigma(N) u_2/u & X_0 (N) - i X_\sigma(N) u_3/u
\end{pmatrix}.
\end{eqnarray}
Now, we can rewrite transmission probabilities in terms of $N$ for both cases
\begin{eqnarray}
T_B(N) = |t_B|^2 &=& \left|X_0 (N) - X_\sigma(N) \tan \theta_1^B \dfrac{u_3}{u} + i \dfrac{X_\sigma(N)}{\cos \theta_1^B} \dfrac{u_2}{u} \right|^{-2}, \nonumber\\
T_E(N) = |t_E|^2 &=& \left|X_0 (N) - X_\sigma(N) \tan \theta_1^E \dfrac{u_3}{u} - i \dfrac{X_\sigma(N)}{\cos \theta_1^E} \dfrac{u_1}{u} \right|^{-2}.
\end{eqnarray}
We can see that the number $N$ of barriers plays the role of a factor in the phase of the trigonometric functions, making the transmission probability $T$ oscillates more rapidly with respect to the incident angle $\alpha$ (keep in mind that all $u,u_1,u_2,u_3,\theta_1^B,\theta_1^E$ depend on $\alpha$). The obvious consequence is that when $N$ is doubled, the number of transparent peaks is roughly doubled as well, as observed in the Figure \ref{trans-multi}.

\end{document}